\documentclass[twocolumn]{journal}

\usepackage{graphicx}
\usepackage{cite}
\usepackage{amsmath,amssymb,amsfonts}
\usepackage{subfigure}
\usepackage{textcomp}
\usepackage{xcolor}

\usepackage{enumerate}

\usepackage{tcolorbox}

\usepackage{algpseudocode}
\usepackage{algorithmicx,algorithm}

\usepackage{pgf}
\usepackage{tikz}
\usetikzlibrary{positioning, arrows.meta, automata}
\usetikzlibrary{calc}
\usetikzlibrary{patterns}
\usepackage[latin1]{inputenc}

\newtheorem{theorem}{Theorem}

\newtheorem{example}{Example}

\newtheorem{definition}{Definition}
\newtheorem{proposition}{Proposition}

\begin{document}

\def\XY#1{{\textcolor{red}{ {\bf XY:} #1}}}

\begin{frontmatter}

\title{On Epistemic Properties in Discrete-Event Systems: \\ 
A Uniform Framework and Its Applications \thanksref{footnoteinfo}} 

\thanks[footnoteinfo]{This work was supported by the National Natural Science Foundation of China (62173226,62061136004). 
}

\author[First]{Bohan Cui} 
\author[Second]{Ziyue Ma}
\author[First]{Shaoyuan Li}
\author[First]{Xiang Yin}

\address[First]{Department of Automation, Shanghai Jiao Tong University, Shanghai, 200240, China.\\ (E-mail: \{bohan\_cui, syli, yinxiang\}@sjtu.edu.cn).}
\address[Second]{School of Electro-Mechanical Engineering, Xidian University, Xi'an 710071, China.\\
(E-mail: maziyue@xidian.edu.cn).}


\begin{abstract}
In this paper, we investigate the property verification problem for partially-observed DES from a new perspective. 
Specifically, we consider the problem setting where the system is observed by two agents independently, each with its own observation.  
The purpose of the first agent, referred to as the \emph{low-level observer}, is to infer the actual behavior of the system, while the second, referred to as the \emph{high-level observer}, aims to infer the knowledge of Agent 1 regarding the system.  We present  a general notion called the \emph{epistemic property} capturing the inference from the high-level observer to the low-level observer. 
A typical instance of this  definition is the notion of high-order opacity, 
which specifies that the intruder does not know that the system knows some critical information. 
This  formalization is very general and supports any user-defined information-state-based knowledge between the two observers.  We demonstrate how the general definition of epistemic properties can be applied in different problem settings such as  information leakage diagnosis or tactical cooperation without explicit communications. 
Finally, we   provide a systematic approach for the verification of epistemic properties. Particularly, we identify some fragments of  epistemic properties that can be verified more efficiently. 
\end{abstract}

\end{frontmatter}

\section{Introduction}

\subsection{Backgrounds and Motivations}

This paper focuses on the property analysis of discrete event systems (DES), an important class of dynamical systems characterized by discrete-state spaces and event-triggered dynamics.  
Many man-made complex engineering systems, such as manufacturing systems, industrial control systems, and logistic systems, can be effectively modeled as DES due to their inherently discrete nature in state transitions. 
Furthermore, DESs also provide a fundamental formal model for specifying the high-level behaviors of cyber-physical systems that involve both discrete and continuous spaces \cite{cassandras2008introduction}.

In many applications, DESs are partially observed in the sense that the system user, or more generally an external observer, cannot directly access the state information due to, for example, the lack of sensors or partial information release \cite{hadjicostis2020estimation,basilio2021analysis,liu2022secure}. Instead, one can only access the \emph{information-flow} generated by the system, which is a projection of the system behavior.  In the context of partially observed DES, one of the most fundamental questions is whether or not the information-flow contains sufficient information for  some decision problem. This leads to the observational property analysis problem, which has been extensively studied in the literature, such as diagnosability in the context of fault detection or opacity in the context of security analysis.

\begin{figure}[tp]
	\centering
	\includegraphics[width=1\linewidth]{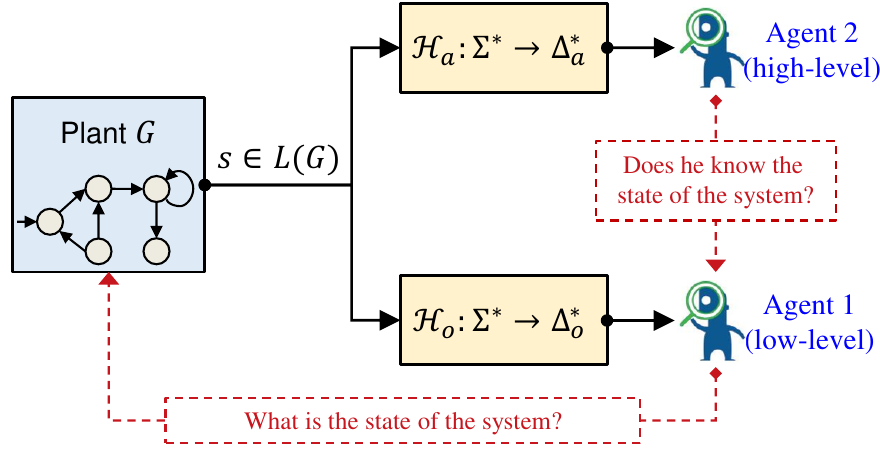}
	\caption{Conceptual illustration of the inferring of the knowledge.}
	\label{fig:illu}
\end{figure}

Most existing works on property analysis of partially-observed DES consider the scenario where a single agent aims to infer the behavior of the system based on its observations. For example, both diagnosability and opacity belong to this setting, although the single observation agent has different roles and purposes.
However, as shown in Figure~\ref{fig:illu}, in some applications, the system may be observed independently by multiple agents with different observations. In such a scenario, one may not only be interested in inferring the system's behavior directly but may also be interested in inferring the other agents' knowledge of the system indirectly.

To be more specific, let us consider a secure state detection setting, where, for utility purposes, Agent 1, who is the system user, wants to determine the precise current state of the system. At the same time, he does not want Agent 2, who is a malicious observer, to know that he has determined the state. Therefore, such high-order inferences arise naturally in partially-observed DES with multiple observation agents. In this work, we will refer to such requirements involving high-order inferences as \emph{epistemic properties}.
Yet, most existing works still focus on the single-agent setting and lack systematic definitions and analysis techniques for epistemic properties.

\subsection{Our  Contributions}
In this paper, we investigate the property verification problem for partially-observed DES from a new perspective. Specifically, we consider the scenario where the system is observed by two agents independently, each with its own observation. The purpose of Agent 1, referred to as the \emph{low-level observer}, is to infer the actual behavior of the system, such as whether or not it can distinguish some critical states. On the other hand, Agent 2, referred to as the \emph{high-level observer}, aims to infer the knowledge of Agent 1 regarding the system. Based on this two-agent model, the main contributions of this paper are summarized as follows:\vspace{-6pt}
\begin{itemize}
\item
First, we formally define a general notion of epistemic property capturing the inference from the high-level observer to the low-level observer. The definition is parameterized by the string quantifier, low-level requirements, and high-level requirements. This general formalization supports any user-defined information-state-based knowledge between the two observers, such as ``You don't know that I know" or ``I know that you don't know".\medskip
\item
We demonstrate how the general definition of epistemic properties can be applied in different problem settings such as knowledge security, attack invisibility, insecure awareness, and tacit cooperation. For example, the knowledge security problem generalizes the standard opacity analysis from protecting the release of some secret state to the release of some critical knowledge.
\medskip
\item
Then we develop a systematic approach for the verification of epistemic properties. Specifically, we first provide a general approach with doubly-exponential complexity to verify an arbitrary epistemic property whose low-level knowledge is information-state based. 
For specific types of epistemic properties, we show the verification complexity can be further reduced to a single exponential.\medskip
\item
Finally, we discuss how the proposed notion of epistemic property can be further generalized when delayed information is involved rather than instant knowledge. This generalization allows us to extend the standard diagnosability analysis from diagnosing the occurrence of faults to detecting certain critical information leakages. 
\end{itemize}


\subsection{Related Works}

\textbf{Analysis of Partially-Observed DES. }
The exploration of observational property verification dates back to the early investigations of supervisory control for partially-observed DESs, where the notion of observability was investigated \cite{lin1988observability}. Observability is essentially a special instance of distinguishability \cite{wang2007algorithm,sears2014computing}, requiring that one should always have sufficient information to resolve control conflicts. Another similar notion is detectability, which requires determining the precise state of the system \cite{shu2012detectability}. 
In the context of fault diagnosis (resp., prognosis, the notions of diagnosability \cite{sampath1995diagnosability, lin1994diagnosability,carvalho2012robust,takai2021general,ma2024verification} (resp., predictability  \cite{genc2009predictability,ran2022prognosability,chen2022stochastic}) have also been widely investigated, specifying that any fault event can always be determined after (resp., before) its occurrence.
In the context of information-flow analysis, various notions of opacity have been proposed to capture the plausible deniability of the system's security against a passive intruder \cite{lin2011opacity,saboori2011inf,tong2017verification,yin2019infinite,ma2021verification,tong2022verification,balun2023verifying}. 
However, all these properties only consider the setting of a single observation agent.

\textbf{Decentralized Decision-Making in DES. }
The issue of knowledge reasoning over multiple observers also arises in decentralized decision-making for DES \cite{ricker2000know,ricker2007knowledge,ritsuka2022epistemic,ritsuka2023you,zhang2023unified,ritsuka2024uniform}. For example, in the context of decentralized supervisory control, the notions of conditional co-observability \cite{yoo2004decentralized} and inference-observability \cite{takai2008synthesis} capture one agent's knowledge regarding the other agent's knowledge.
However, these works mainly consider a collaborative setting, where all agents need to complete a specific global task. In contrast, we consider a more generic and parameterized notion of epistemic property, allowing for specifying security-type requirements in antagonistic settings, which is not addressed in existing works on decentralized decision-making.

\textbf{Hyper-Properties and Epistemic Logic. }
Our work is conceptually related to the notion of hyper-properties \cite{beutner2023second} in the context of model checking, and the notion of dynamic epistemic logic \cite{van2007dynamic} in the context of knowledge theory. Both of these frameworks allow descriptions such as ``you know that I know". However, in these works, the knowledge relationships are typically interpreted using Kripke structures rather than partially-observed DES. 
For example, recent work has shown that there is a modeling gap between partially-observed DES and Kripke structures, even for standard observational properties \cite{zhao2024unified}. Furthermore, the verification of second-order hyper-properties is undecidable in general. Our work here essentially provides a class of epistemic properties that can be effectively verified directly based on partially-observed DES models.

\textbf{Our Preliminary Works. }
In our preliminary works \cite{cui2022you,cui2024}, we have studied two specific types of epistemic properties. In \cite{cui2022you}, we proposed the notion of high-order opacity, which captures the requirement that the high-level observer should never be certain that the low-level observer has detected some fact. In \cite{cui2024}, we introduced the notion of epistemic diagnosability, which ensures that the high-level observer should always be able to detect critical information leakage at the low level. However, these two works follow a case-by-case approach to specifying properties for specific scenarios. This journal version provides a much more general framework that not only subsumes our preliminary works \cite{cui2022you,cui2024}, but also supports more user-defined epistemic properties for different applications.

\subsection{Organization}
The rest of this paper is organized as follows. 
Section~\ref{sec-pre} provides some basic preliminaries. 
In Section~\ref{sec-mot}, we present a motivating example and discuss our main idea before we formally present our results. 
Then, we introduce the description of \emph{external knowledge} and outline the uniform notion of \emph{epistemic properties} in Section~\ref{sec-notion}. 
Additionally, we provide several applications of our framework in different scenarios in Section~\ref{sec-appli}.
Then we provide the verification procedure and present several approaches with lower complexity in Section~\ref{sec-verify}.
In Section~\ref{sec-delay}, we discuss the extension of epistemic properties to the case of delayed knowledge.
Finally, we conclude the paper and discuss future directions in Section~\ref{sec-con}.

\section{Preliminaries}\label{sec-pre}

Let $\Sigma$ be a finite set of events.  
A string is a finite sequence of events and $\Sigma^*$ denotes the set of all strings over $\Sigma$ including the empty string $\epsilon$. 
For any string $s\in \Sigma^*$, $|s|$ denotes the length of $s$ with $|\epsilon|=0$. 
A language $L\subseteq\Sigma^*$ is a set of strings. 
For any string $s\in L$, we denote by $L/s$ the post-language of $s$ in $L$, i.e., $L/ s:=\{w\in \Sigma^*: sw\in L\}$. 
Also, we denote by $\overline{L}$ the prefix-closure of language $L$, i.e., $\overline{L}=\{s\in\Sigma^*:\exists w\in\Sigma^* \text{ s.t. }  sw\in L\}$.

We consider a DES modeled by a  deterministic finite-state automaton (DFA)\vspace{-3pt} 
\[
G=(X,\Sigma,\delta,x_0),  \vspace{-3pt} 
\]
where $X$ is a finite set of states,  
$\Sigma$ is a finite set of events,  
$\delta:X \times \Sigma \to X$ is the partial transition function such that for any $x,x'\in X$, $\sigma\in\Sigma$, $x'=\delta(x,\sigma)$ means that there exists a transition from state $x$ to state $x'$ via event $\sigma$, 
and  $x_0\in X$ is the initial state. 
The transition function is also extended to $\delta: X\times \Sigma^* \to X$ recursively by: 
 (i)  for any $x\in X$, $\delta(x,\epsilon)=x$; and 
 (ii) for any $x\in X, s\in \Sigma^*, \sigma \in \Sigma$, we have $\delta(x,s\sigma)= \delta( \delta(x,s) ,\sigma)$. 
The set of all strings  generated by $G$ starting from state $x\in X$ is defined as $\mathcal{L}(G,x)=\{s \in \Sigma^*:\delta(x,s)!\}$, where ``$!$" means ``is defined". 
The set of all strings generated by $G$ is defined as $\mathcal{L}(G):=\mathcal{L}(G,x_0)$. For any $s\in\mathcal{L}(G)$,  we write $\delta(x_0,s)$ simply as $\delta(s)$. 
For the sake of simplicity, we assume that system $G$ is live, i.e., for any $x\in X$, there exists $\sigma\in \Sigma$ such that $\delta(x,\sigma)!$.

When the system is partially observed, the occurrence of each
event is imperfectly observed (either by the system user or an external observer) through an observation mapping defined as follows \vspace{-3pt} 
\[
\mathcal{H}:\Sigma\to \Delta\dot{\cup}\{\epsilon\}, \vspace{-3pt} 
\]
where $\Delta$ is a new set of observation symbols. 
That is, we
observe $\mathcal{H}(\sigma)$ upon the occurrence of event $\sigma\in\Sigma$. 
We say
event $\sigma\in \Sigma$ is observable if $\mathcal{H}(\sigma)\in \Delta$ and unobservable if $\mathcal{H}(\sigma)= \epsilon$. 
The observation function is extended to $\mathcal{H}:\Sigma^*\to\Delta^*$ by: for any $s\in\Sigma^*$, $\mathcal{H}(s)$ is obtained by replacing each event $\sigma$ in string $s$ as $\mathcal{H}(\sigma)$. We denote by $\mathcal{H}^{-1}:\Delta^* \to 2^{\Sigma^*}$ the inverse function, i.e., for any $\alpha\in \Delta^*$, we have $\mathcal{H}^{-1}(\alpha)=\{s\in \Sigma^*: \mathcal{H}(s)=\alpha\}$. The observation function is also extended to $\mathcal{H}: 2^{\Sigma^*}\to 2^{\Delta^*}$ by: for any $L\subseteq \Sigma^*$, $\mathcal{H}(L)=\{\mathcal{H}(s)\in \Delta^*: s\in L\}$. We also extend the inverse function to $\mathcal{H}^{-1}:2^{\Delta^*}\to 2^{\Sigma^*}$ analogously.

In partially observed DES,  one can estimate the states the system is in  based on the online observation  and the system model $G$.
Formally,  let $\alpha\in \mathcal{H}(\mathcal{L}(G))$ be an observed string.  The \emph{current-state estimate} upon the observation of $\alpha$  is the set of  all possible states the system can be in currently when $\alpha$ is observed, i.e., \vspace{-3pt} 
        \[
        \hat{X}(\alpha)= 
        \left\{
            \delta(s)\in X:  
         \exists s\in \mathcal{L}(G)\text{ s.t. } \mathcal{H}(s)=\alpha 
        \right\}. 
        \]

\section{Motivating Example}\label{sec-mot}
In the analysis of partially-observed DES, most existing works consider the setting where the system's behavior is observed by a single agent. 
However, in some applications, there may be multiple observation agents monitoring the system's behavior independently, each with its own mapping. In such cases, one agent may not only be interested in what it knows about the system but may also be interested in \emph{what other agents know}. We refer to such observational properties involving the knowledge of multiple agents as \emph{high-order properties}.
Before formally formulating the high-order property, we use the following motivating example to illustrate our idea.

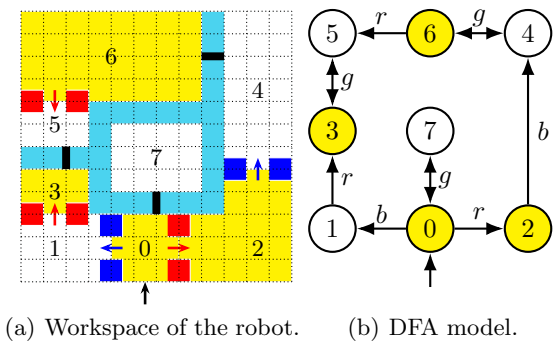
\begin{figure}
  \centering
    \subfigure[Workspace of the robot.\label{motiexample-1}]{\centering
       \definecolor{mygray}{RGB}{192,192,192}
    \definecolor{mycyran}{RGB}{75,211,239}
    \begin{tikzpicture}

    	\draw[fill=mycyran, draw = white] (4.4, 0) -- (8, 0) -- (8, 3.6) -- (4.4, 3.6);
    	
    	\draw[fill=yellow, draw = white] (5.6, 0) -- (6.5, 0) -- (6.5, 0.9) -- (5.6, 0.9);
    	
    	\draw[fill=white, draw = white] (5.6,1.2) -- (6.8,1.2) -- (6.8, 2.1) -- (5.6, 2.1);

    	\draw[fill=yellow, draw = white] (6.5,0) -- (8,0) -- (8, 0.9) -- (6.5, 0.9);
    	\draw[fill=yellow, draw = white] (7.1,0) -- (8,0) -- (8, 1.5) -- (7.1, 1.5);

    	\draw[fill=white, draw = white] (7.1,1.5) -- (8,1.5) -- (8, 3.6) -- (7.1, 3.6);
    	
    	\draw[fill=yellow, draw = white] (4.4, 2.4) -- (6.8, 2.4) -- (6.8, 3.6) -- (4.4, 3.6);
    	
    	\draw[fill=white, draw = white] (4.4,1.8) -- (5.3,1.8) -- (5.3, 2.4) -- (4.4, 2.4);
    	
    	\draw[fill=yellow, draw = white] (4.4,0.9) -- (5.3,0.9) -- (5.3, 1.5) -- (4.4, 1.5);    
    	
    	\draw[fill=white, draw = white] (4.4,0) -- (5.6,0) -- (5.6, 0.9) -- (4.4, 0.9); 
    	
    	\begin{scope}[thick, black, line width=0.35mm, arrows = {-Stealth[scale=0.6]}]
    		\draw (6.05,-0.3) -- (6.05,0);
    	\end{scope}
    	\draw[line width=3pt, black] (6.2, 0.9)--(6.2, 1.2);  
    	\draw[line width=3pt, black] (6.8, 3.0)--(7.1, 3.0);
    	\draw[line width=3pt, black] (5.0, 1.5)--(5.0, 1.8);
    	\draw[fill=red, draw = white] (6.35,0) -- (6.65,0) -- (6.65, 0.3) -- (6.35, 0.3);
    	\draw[fill=red, draw = white] (6.35,0.6) -- (6.65,0.6) -- (6.65, 0.9) -- (6.35, 0.9);
    	\begin{scope}[thick, red, line width=0.35mm, arrows = {-Stealth[scale=0.6]}]
    		\draw (6.35,0.45) -- (6.65,0.45);
    	\end{scope}
    	
    	\draw[fill=red, draw = white] (4.4,2.25) -- (4.7,2.25) -- (4.7, 2.55) -- (4.4, 2.55);
    	\draw[fill=red, draw = white] (5.0,2.25) -- (5.3,2.25) -- (5.3, 2.55) -- (5.0, 2.55);
    	\begin{scope}[thick, red, line width=0.35mm, arrows = {-Stealth[scale=0.6]}]
    		\draw (4.85,2.55) -- (4.85,2.25);
    	\end{scope}

    	\draw[fill=red, draw = white] (4.4,0.75) -- (4.7,0.75) -- (4.7, 1.05) -- (4.4, 1.05);
    	\draw[fill=red, draw = white] (5.0,0.75) -- (5.3,0.75) -- (5.3, 1.05) -- (5.0, 1.05);    	
    	\begin{scope}[thick, red, line width=0.35mm, arrows = {-Stealth[scale=0.6]}]
    		\draw (4.85,0.75) -- (4.85,1.05);
    	\end{scope}
    	
    	\draw[fill=blue, draw = white] (7.1,1.35) -- (7.4,1.35) -- (7.4, 1.65) -- (7.1, 1.65);
    	\draw[fill=blue, draw = white] (7.7,1.35) -- (8.0,1.35) -- (8.0, 1.65) -- (7.7, 1.65);
    	\begin{scope}[thick, blue, line width=0.35mm, arrows = {-Stealth[scale=0.6]}]
    		\draw (7.55,1.35) -- (7.55,1.65);
    	\end{scope}
    	\draw[fill=blue, draw = white] (5.45,0) -- (5.75,0) -- (5.75, 0.3) -- (5.45, 0.3);
    	\draw[fill=blue, draw = white] (5.45,0.6) -- (5.75,0.6) -- (5.75, 0.9) -- (5.45, 0.9);
    	\begin{scope}[thick, blue, line width=0.35mm, arrows = {-Stealth[scale=0.6]}]
    		\draw (5.75,0.45) -- (5.45,0.45);
    	\end{scope}
    	\foreach \x in {4.4,4.7,...,8.3}
    	\draw[densely dotted] (\x, 0)--(\x, 3.6);
    	\foreach \y in {0,0.3,...,3.9}
    	\draw[densely dotted] (4.4, \y)--(8, \y);
    	
    	\node[] [xshift=6.05cm, yshift=0.45cm] (0label) {\footnotesize$0$};
    	\node[] [xshift=4.85cm, yshift=0.45cm] (1label) {\footnotesize$1$};
    	\node[] [xshift=7.55cm, yshift=0.45cm] (2label) {\footnotesize$2$};
    	\node[] [xshift=4.85cm, yshift=1.2cm] (3label) {\footnotesize$3$};
    	\node[] [xshift=7.55cm, yshift=2.55cm] (4label) {\footnotesize$4$};
    	\node[] [xshift=4.85cm, yshift=2.1cm] (5label) {\footnotesize$5$};
    	\node[] [xshift=5.6cm, yshift=3cm] (6label) {\footnotesize$6$};
    	\node[] [xshift=6.2cm, yshift=1.65cm] (7label) {\footnotesize$7$};

    \end{tikzpicture}}
	\subfigure[DFA model.\label{motiexample-2}]{\centering
   \begin{tikzpicture}[->,>={Latex}, thick, initial text={}, node distance=1.3cm, initial where=below, thick, base node/.style={circle, draw, minimum size=6mm}]  
   \node[state, initial, base node, fill=yellow] (0) {$0$};
   \node[state, base node, ] (1) [left of=0] {$1$};
   \node[state, base node, fill=yellow] (2) [right of=0] {$2$};
   \node[state, base node, fill=yellow] (3) [above of=1] {$3$};
   \node[state, base node, ] (5) [above of=3] {$5$};
   \node[state, base node, fill=yellow] (6) [right of=5] {$6$};
   \node[state, base node, ] (4) [right of=6] {$4$};
   \node[state, base node, ] (7) [above of=0] {$7$};
   
   \path[<->]
   (0) edge node [xshift=0.2cm] {$g$} (7)
   (3) edge node [xshift=0.2cm] {$g$} (5)
   (4) edge node [yshift=0.2cm] {$g$} (6);
   \path[->]
   (0) edge node [yshift=0.2cm] {$r$} (2)
   (0) edge node [yshift=0.2cm] {$b$} (1)   
   (6) edge node [yshift=0.2cm] {$r$} (5)
   (1) edge node [xshift=0.2cm] {$r$} (3)
   (2) edge node [xshift=0.2cm] {$b$} (4);
   \end{tikzpicture}}  \\
  \caption{A motivating example with $\Sigma_{o}=\{r,g\}$, $\Sigma_{a}=\{b,g\}$.}
	\label{motiexample}
\end{figure}

We consider robot path planning, where a robot moves in a workspace with rivers, bridges, and  checkpoints as shown in Figure~\ref{motiexample-1}. 
Suppose that the robot can cross the bridges (denoted by black lines) via both directions, but the checkpoints (denoted by red, and blue blocks) are only one-way whose directions are specified by arrows in the figure. 
The mobility of the robot can be modeled as DFA $G$ shown in Figure~\ref{motiexample-2}, where states correspond to regions in the workspace and 
events $b,r$ and $g$ corresponds to  ``passing a blue checkpoint", ``passing a red checkpoint"  and ``crossing the bridge", respectively. 

\emph{Observers' Model: } 
Now let us assume that there are two agents, a user, and an intruder, that observe the behavior of the robot passively and independently\vspace{-6pt}
\begin{itemize}
    \item 
    System User:    
    Suppose that the system user of the robot is a central station seeking to send messages to the robot. However, communication signals are limited to specific service regions and communication signals are not available in areas marked in yellow in Figure~\ref{motiexample-1}, corresponding to yellow states ${0,2,3,6}$ in the DFA model. To ensure reliable reception of messages, the central station relies on sensors positioned at bridges and red checkpoints, enabling it to detect events $g$ and $r$. Specifically, to guarantee successful transmission \emph{with certainty}, the central station will only send messages to the robot when it confirms that the robot is currently not in yellow regions.
    \medskip
    \item 
    Intruder:
    Meanwhile, we assume that 
    there is an intruder that has sensors positioned at bridges and blue checkpoints, allowing it to observe the occurrences of events $g$ and $b$.  By knowing the strategy of the central station, the intruder endeavors to breach the communication channel between the central station and the robot. Therefore, it will successfully intercept the transmitted message when it knows \emph{for sure} that the central station knows for sure that the robot is at a service region.\vspace{-6pt}
\end{itemize}

We assume that the \emph{robot's task} is achieving at least one successful communication, i.e., there is a moment when the user confirms the robot is not in yellow regions, while simultaneously ensuring that the intruder remains unaware of the user's knowledge.

\emph{Analysis:}
In order to establish communications with the central station,  the robot may choose path $0\overset{g}{\rightarrow}7$, which is the shortest path to reach a service region. The central station will observe event $g$ and upon which it knows for sure that the robot is indeed in a service region $7$. Hence, it will send messages to the robot. However, at the same time, the intruder will also observe event $g$, upon which it knows for sure that the central station will send messages, which makes the system not secure. 

To establish secure communication, the robot can choose the path 
$0\overset{r}{\rightarrow}2\overset{b}{\rightarrow}4\overset{g}{\rightarrow}6\overset{r}{\rightarrow}5$ to go to region $5$. 
Along this path, the user will observe $rgr$ and it knows for sure that the robot is in a service region. 
Meanwhile, the intruder will observe $bg$ and it may think that the robot may have chosen path $0\overset{b}{\rightarrow}1\overset{r}{\rightarrow}3\overset{g}{\rightarrow}5$. If this is the case, then the central station will observe $rg$ 
and it cannot distinguish if the robot is at state $5$ or state $6$.  
Therefore, along this path, the central station will know for sure that the robot is in a service region, while the intruder does not know that the central station knows that. This means that the communication in this scenario is secure.

Intuitively, the above knowledge requirement can be described as ``\emph{you don't know that I know}". This requirement is related to security considerations when more than one observers are involved.
 In fact, in different applications, one may further impose knowledge requirements such as ``\emph{you always know that I know}" or ``\emph{you never know that I don't know}". The main purpose of this work is to provide a unified formal definition and verification algorithms for such high-order properties.

\section{General Definition of Epistemic Properties}\label{sec-notion}

\subsection{Two-Agent Knowledge Model}
To formally describe the epistemic properties that arise when there are multiple observation sites involved, 
here we consider two distinct agents as follows:\vspace{-6pt} 
\begin{itemize}
    \item 
    The first agent observes the behavior of the system 
    and its purpose is to obtain certain knowledge regarding the system itself; 
    we denote by  $\mathcal{H}_o:\Sigma\to \Delta_o\dot{\cup}\{\epsilon\}$ the observation function of the first agent; and\medskip
    \item 
    The second agent also observes the behavior of the system but its purpose is to infer the knowledge of the first agent by its own observation indirectly. 
    We denote by  $\mathcal{H}_a:\Sigma\to \Delta_a\dot{\cup}\{\epsilon\}$ the observation function of the second agent.\vspace{-6pt}
\end{itemize}

Hereafter, we will refer to the first agent as the \emph{low-level observer} and the second agent as the \emph{high-level observer}. This terminology simply aims to characterize the fact that the high-level observer seeks to infer the knowledge of the low-level observer. It does not imply any inherent relationship between their observation functions.
Specifically, we denote the sets of observable events for the low-level observer and the high-level observer as $\Sigma_o$ and $\Sigma_a$, respectively, with $\Sigma_{uo} := \Sigma \setminus \Sigma_o$ and $\Sigma_{ua} := \Sigma \setminus \Sigma_a$.
In general, $\Sigma_a$ and $\Sigma_o$ can be incomparable. 
Also, we remark that
the system user can be either the low-level observer or the high-level observer, depending on the specific problem scenario, which we will discuss in more detail later.

\subsection{Knowledge of the Low-Level Observer}
As we mentioned, the low-level observer aims to obtain certain knowledge of the system, e.g., for users' decision-making such as control or diagnosis and, for system intruders' secret-stealing such as destroying the opacity or anonymity of systems.
In general, such knowledge can be formally described by a predicate\vspace{-3pt} 
\begin{equation}
\texttt{Kw}_o: \mathcal{H}_o(\mathcal{L}(G)) \to \{\texttt{T},\texttt{F}\}\vspace{-3pt}
\end{equation}
such that  for any $\alpha\in \mathcal{H}_o(\mathcal{L}(G))$, $\texttt{Kw}_o(\alpha)=\texttt{T}$ means that 
some knowledge of interest holds based on observation $\alpha$. 
Note that, the above-defined knowledge is language-based, which may require infinite memory to realize. In this paper, we consider the following \emph{information-state-based} knowledge, which is general enough to capture a large class of practical requirements. 

\begin{definition}[IS-Based Knowledge]\label{def-IS}\upshape 
An information-state-based (IS-based)  knowledge is a predicate on the state estimate of the system, i.e., \vspace{-3pt}
\begin{equation}
\texttt{Kw}_o: 2^X \to \{\texttt{T},\texttt{F}\}.\vspace{-3pt}
\end{equation}
For any $\alpha\in \Delta^*_o$,  
we define 
$\texttt{Kw}_o(\alpha)\!=\!\texttt{T}$ iff 
$\texttt{Kw}_o(\hat{X}_o(\alpha))\!=\!\texttt{T}$, 
where $\hat{X}_o(\cdot)$ is the state estimate w.r.t.\ $\mathcal{H}_o$.
\end{definition}

Intuitively, IS-based knowledge says that the lower-level observer can determine its knowledge solely based on its state estimate without further memory. 
In this paper, we will mainly explore a widely used IS-based knowledge  called \emph{distinguishability}. 
Specifically, let\vspace{-3pt}
    \[
    T_{\texttt{spec}}\subseteq X\times X\vspace{-3pt}
    \] 
be a set of state pairs one needs to distinguish. 
Then the state distinguishability can be specified by IS-based knowledge, 
denoted by  $\texttt{Dis}_o^{T_\texttt{spec}}: 2^X\to \{\texttt{T},\texttt{F}\}$, 
such that for any $q\in 2^X$, we have\vspace{-3pt} 
\begin{equation}
    \texttt{Dis}_o^{T_\texttt{spec}}(q)=\texttt{T}
    \Leftrightarrow
    (q\times q)\cap T_{\texttt{spec}}=\emptyset.\vspace{-3pt}
\end{equation}
In what follows, will omit the subscript $T_\texttt{spec}$ and write it as $\texttt{Dis}_o$ when it is clear from the context. 
The distinguishability  can be understood as follows: if  $\texttt{Dis}_o(\alpha)=\texttt{T}$, 
then there do not exist two strings  with the same observation  $\alpha$ such that they lead to a pair of two states in  $T_{\texttt{spec}}$.

The definition of distinguishability is general enough for many practical requirements. Here are two examples:\vspace{-6pt}
\begin{itemize}
    \item 
    In the context of fault diagnosis, 
let $X=X_N\dot{\cup}X_F$, where $X_N$ is the set of normal states and $X_F$ is the set of fault states. 
By defining $T_\texttt{spec}=X_N\times X_F$,  then
$\texttt{Dis}_o(q)=\texttt{T}$ means that 
the low-level observer is sure about the faulty status of the system, i.e., 
either $q\subseteq X_N$ or $q\subseteq X_F$. \medskip
\item 
In the context of security analysis, 
let $X=X_S\dot{\cup}X_{NS}$, where $X_S$ is the set of secret states and $X_{NS}$ is the set of non-secret  states. 
By defining $T_\texttt{spec}=X_{NS}\times X_{NS}$, then
$\texttt{Dis}_o(q)=\texttt{T}$ means that 
the low-level observer knows for sure that it is currently at a secret state, i.e., $q\subseteq X_S$. 
\end{itemize}




\subsection{Knowledge of the High-Level Observer}
On the other hand, the  high-level observer independently makes observations but aims to infer the knowledge of the low-level observer. 
Formally, we assume that the high-level observer has the following capabilities: \vspace{-6pt}
\begin{itemize}
    \item
    It knows the DFA model  $G$ of the system;\medskip
    \item
    It can observe the occurrence of each event in $\Sigma_a$ generated online through function $\mathcal{H}_a$; \medskip
    \item
    It knows that the low-level observer can observe the occurrence of each event in $\Sigma_o$ online through $\mathcal{H}_o$, but it cannot observe the occurrences of events $\Sigma_o \setminus \Sigma_a$. 
\end{itemize}

The basic idea of epistemic properties still follows the essence of observational properties, i.e., being able to know (or not know) some ``critical behaviors". Differently, in the setting of epistemic properties, ``critical behaviors" are not captured by states or events of the system itself as most observational properties did, 
and they are described by the knowledge of the low-level observer. 
To this end, one of the key elements of the notion of epistemic properties is \emph{the estimate of the knowledge}.
Formally, given partially observed system $G$ with observation function $\mathcal{H}_o$ and $\mathcal{H}_a$, the intruder's estimate of the user's knowledge can be modeled as a predicate on the intruder's observations
\[
\widehat{\texttt{Kw}}_{ao}: \mathcal{H}_a(\mathcal{L}(G))\to \{\texttt{Y},\texttt{N},\texttt{U}\}
\]
such that for any $\alpha\in\mathcal{H}_a(\mathcal{L}(G))$, we have
\begin{align} 
\widehat{\texttt{Kw}}_{ao}(\alpha) \!=\!
		\left\{
		\begin{array}{ll}
			\texttt{Y} &  \quad\text{if } \ (\forall\omega\in \mathcal{H}_{ao}^{-1}(\alpha) )[\texttt{Kw}_o(\omega)=\texttt{T}]\\
			\texttt{N} &   \quad\text{if }  \    (\forall\omega\in \mathcal{H}_{ao}^{-1}(\alpha) )[\texttt{Kw}_o(\omega)=\texttt{F}]\\
			\texttt{U} &  \quad \text{otherwise}
		\end{array}.
		\right.   
\end{align} 
where $\mathcal{H}_{ao}^{-1}(\alpha):=\mathcal{H}_o(  \mathcal{H}_a^{-1}(\alpha) \cap \mathcal{L}(G) )$.

The above definition is intuitively explained as follows. 
Upon the occurrence of string $s\in\mathcal{L}(G)$,  the low-level observer observes $\mathcal{H}_o(s)$ while the high-level observer observes $\mathcal{H}_a(s)$. 
As we discussed before, the knowledge of the low-level observer is completely determined by $\mathcal{H}_o(s)$.
However, the high-level observer does not know this knowledge perfectly.  From its point of view, the low-level observer may have observed  any string in $ \mathcal{H}_o(  \mathcal{H}_a^{-1}(\mathcal{H}_a(s)) \cap \mathcal{L}(G) ) $. If for any string $\alpha$ in this set, we have $\texttt{Kw}_o(\alpha)=\texttt{T}$ (resp.\ $\texttt{F}$), 
then the high-level observer knows for sure that predicate $\texttt{Kw}_o$ does (resp.\  does not) hold for the low-level observer, i.e., 
$\widehat{\texttt{Kw}}_{ao}(\mathcal{H}_a(s))=\texttt{Y}$ ($\texttt{F}$). Otherwise, $\widehat{\texttt{Kw}}_{ao}(\mathcal{H}_a(s))=\texttt{U}$.

Now, we are ready to propose a uniform notion of epistemic properties.

\begin{definition}[Epistemic Properties]\label{def-epis}\upshape
Let  $G$ be the system with two observation functions $\mathcal{H}_o$ and $\mathcal{H}_a$, 
and  $\texttt{Kw}_o$ be an IS-based knowledge. 
Then an  epistemic property is of the following form
\begin{equation}\label{eq:epis-prop}
(\mathbb{Q}  s \in\mathcal{L}(G): \texttt{Kw}_o(\mathcal{H}_o(s))=\mathbb{K}_o) 
        [\widehat{\texttt{Kw}}_{ao}(\mathcal{H}_a(s))=\mathbb{K}_{ao}],
\end{equation}
where 
$\mathbb{Q}\in\{\forall,\exists\}$ is the quantifier, 
$\mathbb{K}_o\in \{\texttt{T},\texttt{F}\}$ is the knowledge of the low-level observer, 
and 
$\mathbb{K}_{ao}\in \{\texttt{Y},\texttt{N},\texttt{U}\}$ is the knowledge of the high-level observer. 
\end{definition}

Therefore, tuple $\langle\mathbb{Q},\texttt{Kw}_o,\mathbb{K}_o,\mathbb{K}_{ao}\rangle$ specifies an epistemic property.
We use notation $G\models\langle\mathbb{Q},\texttt{Kw}_o,\mathbb{K}_o,\mathbb{K}_{ao}\rangle$ 
to denote that system $G$ satisfies the epistemic property  w.r.t.\ these paramters. 
For example, consider an epistemic property specified as the tuple $\langle\forall,\texttt{Dis}_o,\texttt{Y},\texttt{Y}\rangle$.  
This property says that, for any string $s\in\mathcal{L}(G)$ upon whose occurrence the low-level observer can distinguish some pairs of states by observing $\mathcal{H}_o(s)$, the high-level observer can always know for sure that the low-level observer knows the fact by observing string $\mathcal{H}_a(s)$. 
This property is related to the so-called tacit cooperation problem, which will be discussed in more detail later. 


\section{Applications of Epistemic Properties}\label{sec-appli}
In this section, we show many useful requirements can be captured by the general definition of epistemic property with some suitable state-space refinement. 
We will first discuss their motivations and then provide their formal definitions.

\subsection{Knowledge Security}

In the motivating example, we considered the scenario where the user is trying to complete the state disambiguation task while also wanting to prevent the intruder from knowing that the task has been successfully achieved.
In this context, 
the system user is the low-level observer and the intruder is the high-level observer.  
We call this requirement \emph{high-order opacity}, which can be formally defined as follows. 

\begin{definition}[High-Order Opacity]\upshape 
Let $G$ be a system with observation functions $\mathcal{H}_o$ and $\mathcal{H}_a$, 
and $T_\texttt{spec}$ be a specification the low-level observer wants to distinguish. 
Then the system is said to be  \emph{high-order opaque}  (w.r.t.\ $T_\texttt{spec}$, $\mathcal{H}_a$ and $\mathcal{H}_o$) if
\begin{align}\label{eq:ho-opc}
&(\forall s\in \mathcal{L}(G):\texttt{Dis}_o(\mathcal{H}_o(s))=\texttt{T})(\exists t\in \mathcal{L}(G)) \nonumber \\
&  [\texttt{Dis}_o(\mathcal{H}_o(t))\!=\!\texttt{F}\wedge \mathcal{H}_a(s)\!=\!\mathcal{H}_a(t)].\vspace{-6pt}
\end{align}
\end{definition}

Intuitively, this requirement states that for any string  $s$ where the user's knowledge is such that $\texttt{Dis}_o(\mathcal{H}_o(s))=\texttt{T}$, the intruder does not have perfect knowledge of the user's understanding. Specifically, from the intruder's perspective, there exists a possible string $t\in   \mathcal{H}_a^{-1}(\mathcal{H}_a(s)) \cap \mathcal{L}(G)$ under which the user's knowledge is $\texttt{Dis}_o(\mathcal{H}_o(t))=\texttt{F}$. Consequently, the intruder cannot be certain whether the user has successfully distinguished the state pairs. 

Clearly, high-order opacity is an instance of epistemic property, which is formalized as follows.  

\begin{proposition}\upshape
System $G$ is \emph{high-order opaque} (w.r.t.\ $T_\texttt{spec}$, $\mathcal{H}_a$ and $\mathcal{H}_o$) if and only if 
$G\models \langle  \forall, \texttt{Dis}_o,\texttt{T},\texttt{U} \rangle$.\vspace{-12pt}  
\end{proposition}
\begin{pf}
The proof is straightforward. 
To see this, we substitute $\langle  \forall, \texttt{Dis}_o,\texttt{T},\texttt{U} \rangle$ into formula~(\ref{eq:epis-prop}) and get\vspace{-6pt}
\begin{equation}
(\forall  s \in\mathcal{L}(G): \texttt{Dis}_o(\mathcal{H}_o(s))=\texttt{T}) 
        [\widehat{\texttt{Kw}}_{ao}(\mathcal{H}_a(s))=\texttt{U}].\nonumber\vspace{-6pt}
\end{equation}
Then, according to the definition of $\widehat{\texttt{Kw}}_{ao}(\mathcal{H}_a(s))$, we  conclude that there exists $\mathcal{H}_o(t)\in\mathcal{H}_{ao}^{-1}(\mathcal{H}_a(s))$ such that $\texttt{Dis}_o(\mathcal{H}_o(t))=\texttt{F}$. For the proof of the other side, since $\mathcal{H}_o(s)\in\mathcal{H}_{ao}^{-1}(\mathcal{H}_a(s))$ by itself, while $\texttt{Dis}_o(\mathcal{H}_o(s))=\texttt{T}$ and $\texttt{Dis}_o(\mathcal{H}_o(t))=\texttt{F}$, we can conclude that $\widehat{\texttt{Kw}}_{ao}(\mathcal{H}_a(s))=\texttt{U}$.
\hfill $\qed$ 
\end{pf}

\subsection{Invisible Attacker}
We use the scenario of \emph{invisible attacker} to further illustrate the general notion of epistemic property 
with the following two purposes:\vspace{-6pt}
\begin{itemize}
\item 
The contexts of the low-level observer and the high-level observer are problem-dependent; and \medskip
\item 
The IS-based knowledge is sometimes defined on a refined state-space.\vspace{-6pt} 
\end{itemize}
To be more specific, we consider the scenario where the intruder wants to detect  \emph{whether or not the system has passed through some critical state}. Meanwhile, it wants to make this detection undetectable from the system user's point of view, ensuring that the system user remains unaware of the disclosure of these secrets. In this context, the low-level observer is the intruder, while the high-level observer is the system user.

Let $X_S\subseteq X$ be the set of critical states that the intruder wants to detect whether or not the system has passed through. 
We define an IS-based knowledge 
$\texttt{Rev}_o: \mathcal{H}_o(\mathcal{L}(G))\to \{\texttt{T}, \texttt{F}\}$ such that, for any $\alpha\in \mathcal{H}_o(\mathcal{L}(G))$, we have $\texttt{Rev}_o(\alpha)=\texttt{true}$ iff  
\vspace{-6pt}
\begin{align}\label{eq:rev}
    (\forall r\in \mathcal{L}(G): \mathcal{H}_o(r)=\alpha) 
    (\exists w\in \overline{\{r\}})
    [ \delta(w)\in X_S]. \vspace{-6pt}
\end{align}
That is, upon observing $\alpha$, the intruder knows for sure that states in $X_S$ have been visited.
Then the above requirement can be captured by the following definition. 

\begin{definition}[Intrusion Undetectability]\upshape
Let $G$ be a system with observation functions $\mathcal{H}_o$ and $\mathcal{H}_a$, 
and $X_S$ be a set of secret states the intruder wants to detect. 
Then the system is said to be  \emph{intrusion undetectable}  (w.r.t.\ $X_S$, $\mathcal{H}_a$ and $\mathcal{H}_o$) if \vspace{-6pt}
\begin{align} \label{intru-undetect}
&(\forall st\in \mathcal{L}(G):\texttt{Rev}_o(\mathcal{H}_o(s))=\texttt{T}) \nonumber\\
&(\exists r\in \mathcal{L}(G): \mathcal{H}_a(r)=\mathcal{H}_a(st))  \\
&(\forall w\in \overline{\{r\}})[\texttt{Rev}_o(\mathcal{H}_o(w))=\texttt{F}].\nonumber\vspace{-6pt}
\end{align}
\end{definition} 
 
The definition above states the following. 
For any  system  string $s\in \mathcal{L}(G)$ such that $\mathcal{H}_o(s)$ leads to a secret release, and any subsequent  string $t\in \mathcal{L}(G)/s$ occurring after $s$, 
 there must exist another string $r$ in $\mathcal{H}_a^{-1}(\mathcal{H}_a(st)) \cap \mathcal{L}(G)$ such that, for all prefixes $w$ of $r$, $\texttt{Rev}_o(\mathcal{H}_o(w))=\texttt{F}$. 
 This means that the  user cannot conclusively determine from its perspective that $\texttt{Rev}_o$ is already true from the intruder's point of view. 

To formulate intrusion undetectability as an epistemic property, one cannot directly define IS-based knowledge based on its original state space. 
This is because the condition in \eqref{eq:rev} requires additional memory to realize. 
Therefore, we refine the original system $G$ by remembering whether or not states in $X_S$ have been visited. This leads to a new augmented system\vspace{-6pt}
\[
    \tilde{G}=(\tilde{X},\Sigma,\tilde{\delta},\tilde{x}_0),\vspace{-6pt}
\]
where
$\tilde{X}\subseteq X\times \{0,1\}$ 
with the initial state be
$(x_0,0)$	if $x_0\not\in X_S$ and $(x_0,1)$	 otherwise, 
and 
$\tilde{\delta}:\tilde{X}\times \Sigma \to \tilde{X}$ is the transition function defined by:
        for any $\tilde{x}=(x,l)\in \tilde{X}$ and $\sigma\in \Sigma$, we have\vspace{-6pt}  
        \[
        \tilde{\delta}(\tilde{x},\tilde{\sigma})=
        \begin{cases}
        	(\delta(x,\sigma),0)&		\text{if } l=0 \land \delta(x,\sigma)\not\in X_S\\
        	(\delta(x,\sigma),1)&		\text{otherwise}\\
        \end{cases}\vspace{-6pt}  
        \]
 
Now we show that   intrusion undetectability is an instance of epistemic property w.r.t.\ the refined state-space $\tilde{G}$. 

\begin{proposition}\upshape
System $G$ is \emph{intrusion undetectable} (w.r.t.\ $X_S$, $\mathcal{H}_a$ and $\mathcal{H}_o$) if and only if  
$\tilde{G}\models \langle  \forall, \texttt{Dis}_o,\texttt{T},\texttt{U} \rangle$, 
where  $\texttt{Dis}_o$ is defined for specification\vspace{-6pt}
\[
  T_{\texttt{spec}}=(X\times \{0\})\times  (X\times \{0\})\subseteq \tilde{X}\times \tilde{X}.\vspace{-6pt}
\]
\end{proposition}
\begin{pf}
Let's first substitute $\langle \forall, \texttt{Dis}_o,\texttt{T},\texttt{U} \rangle$ into formula~(\ref{eq:epis-prop}), then we can get the following equivalent formula.
\begin{align}
    &(\forall  s \in\mathcal{L}(\tilde{G}): \texttt{Dis}_o(\mathcal{H}_o(s))=\texttt{T}) \nonumber \\
    &(\exists r\in\mathcal{L}(\tilde{G}):\mathcal{H}_a(r)= \mathcal{H}_a(s))
    [\texttt{Dis}_{o}(\mathcal{H}_o(r))=\texttt{F}]\nonumber
\end{align} 
By defining $T_\texttt{spec}$ as the above, for any $\alpha\in\mathcal{H}_o(\mathcal{L}(\tilde{G}))$, we have that $\texttt{Dis}_o(\alpha)=\texttt{T}\Leftrightarrow \hat{\tilde{X}}(\alpha)\subseteq (X\times \{1\})$. Then the above formula can be converted to 
\begin{align}
    &(\forall  s \in\mathcal{L}(\tilde{G}): \hat{\tilde{X}}_o(\mathcal{H}_o(s))\subseteq (X\times \{1\})) \nonumber \\
    &(\exists r\in\mathcal{L}(\tilde{G}):\mathcal{H}_a(r)= \mathcal{H}_a(s))    [\hat{\tilde{X}}_o(\mathcal{H}_o(r))\not\subseteq (X\times \{1\})]\nonumber
\end{align}
Note that by construction, $\tilde{G}$ has the following three properties: (i) $\mathcal{L}(G)=\mathcal{L}(\tilde{G})$; (ii) for any $s \in \mathcal{L}(\tilde{G})$ if $\tilde{\delta}(s)\in (X\times \{0\})$, then for any $w\in \overline{\{s\}}$, we have $\delta(w)\not\in X_S$; and (iii) for any $s \in \mathcal{L}(\tilde{G})$ if $\tilde{\delta}(s)\in (X\times \{1\})$, then there must exist $w\in \overline{\{s\}}$ such that $\delta(w)\in X_S$.
Therefore, we can further convert the above formula to 
\begin{align}
    &(\forall  s \in\mathcal{L}(G))(\forall s' \in\mathcal{L}(G):\mathcal{H}_o(s')=\mathcal{H}_o(s)) \nonumber \\
    &(\exists w\!\!\in \overline{\{s'\}}: \delta(w)\in X_S) (\exists r\in\mathcal{L}(G):\mathcal{H}_a(r)= \mathcal{H}_a(s))\nonumber \\
    &(\exists r'\!\!\in\mathcal{L}(G):\mathcal{H}_o(r')=\mathcal{H}_o(r))(\forall w'\in\overline{\{r'\}})[\delta(w')\not\in X_S],\nonumber
\end{align}
which can be further converted to
the formula (\ref{intru-undetect}) according to the definition of $\texttt{Rev}_o$.
\hfill $\qed$ 
\end{pf}


\subsection{Release Awareness}
%
Note that, in the previous two examples, both epistemic properties belong to the same parameter pattern $ \langle  \forall, \texttt{Dis}_o,\texttt{T},\texttt{U} \rangle$. 
The only difference is how $T_{\texttt{spec}}$ is defined for $\texttt{Dis}_o$. 
Here, we provide a different example of epistemic property which adopts a different pattern. 

Specifically, we consider a secret protection scenario similar to the basic setting of opacity, where the system user does not want the intruder to know for certain that it is in a secret state. However, we consider a weaker setting where the system user is allowed to reveal its secret to the intruder, provided that it is fully aware of this information leakage. This weaker definition is motivated by situations where the system user can actively take additional actions to protect itself from being compromised by the intruder once the leakage is detected.

Therefore, in this context, the intruder is the low-level observer 
and the system user is the high-level observer. 
Essentially, the system user aims to \emph{diagnose}  the secret leakage to the intruder. Hence, we term this requirement as the \emph{epistemic  diagnosability}.
 
\begin{definition}[Epistemic  Diagnosability]\upshape\label{def-epis-diag} 
Let $G$ be a system with observation functions $\mathcal{H}_o$ and $\mathcal{H}_a$, 
and $X_S$ be a set of secret states the intruder wants to detect. 
Then system $G$ is said to be \emph{epistemically diagnosable} (w.r.t.\ $X_S$, $\Sigma_a$ and $\Sigma_o$) if\vspace{-6pt} 
\begin{align}\label{eq:eps-diag}
    &(\forall s \in \mathcal{L}(G): \hat{X}_o(\mathcal{H}_o(s))\subseteq X_S)  \\
    &(\forall r\!\in\! \mathcal{L}(G))  [\mathcal{H}_a(s)\!=\!\mathcal{H}_a(r)\Rightarrow \hat{X}_o(\mathcal{H}_o(r))\subseteq X_S ]. \nonumber\vspace{-6pt}
\end{align}
\end{definition}

Intuitively, the above definition implies that once the intruder knows for certain that the system is in a secret state, the system user must also be aware of this. This means that for any other possible strings with the same observation, the secret information is also revealed to the intruder. Note that the diagnosis of leakage is required to be done \emph{instantaneously}. In some applications, this requirement may be relaxed to allow for diagnosing the leakage within a finite number of steps. We will discuss this extension later in Section 7.

Clearly, one can still express epistemic diagnosability as an instance of epistemic property. 

\begin{proposition}\upshape
System $G$ is \emph{epistemically diagnosable} (w.r.t.\ $X_S$, $\mathcal{H}_a$ and $\mathcal{H}_o$) if and only if  
$G\models \langle  \forall, \texttt{Dis}_o,\texttt{T},\texttt{Y} \rangle$, 
where  $\texttt{Dis}_o$ is defined for specification $T_{\texttt{spec}}=(X\setminus X_S)\times(X\setminus X_S)$.\vspace{-6pt}
\end{proposition} 
\begin{pf}
    This proposition can be directly proved by substitute $\langle  \forall, \texttt{Dis}_o,\texttt{T},\texttt{Y} \rangle$ into formula~(\ref{eq:epis-prop}).
\hfill $\qed$ 
\end{pf}

\subsection{Tacit Cooperation}
The previous examples all involve a system user and an intruder in an antagonistic setting. However, epistemic properties can also capture scenarios where two system users are cooperating with each other without direct communication.
 
Similar to the motivating example, we consider a scenario where a robot moves within a workspace and aims to establish communication with a central station. 
In this scenario, we assume the robot has its own partial observation $\mathcal{H}_o$  to infer its location and it will only send a message to the central station when it is certain that it is within a communication region. On the other hand, the central station also has its own partial observation $\mathcal{H}_a$ of the robot, and to conserve energy, it will only activate its radar to receive messages when it is certain that the robot is ready to communicate. Therefore, to avoid missing any messages, it is necessary to ensure that the central station always knows for sure that the robot is certain about its communication locations.

In this context, the robot acts as the low-level observer, while the central station serves as the high-level observer. However, they are in a cooperative setting, as they aim to achieve a common task, i.e., correctly sending and receiving messages, without directly communicating their locations. We term this requirement as \emph{high-order detectability}.

\begin{definition}[High-Order Detectability]\upshape 
Let $G$ be a system with observation functions $\mathcal{H}_o$ and $\mathcal{H}_a$, 
and $T_\texttt{spec}\subseteq X\times X$ be a disambiguation task. 
Then system $G$ is said to be \emph{high-order detectable} (w.r.t.\ $T_\texttt{spec}$, $\Sigma_a$ and $\Sigma_o$) if\vspace{-6pt} 
\begin{align}
&(\forall s\in \mathcal{L}(G))(\forall t\in \mathcal{L}(G)) \nonumber\\
&[\mathcal{H}_{a}(s)=\mathcal{H}_{a}(t)\Rightarrow\texttt{Dis}_{o}(\mathcal{H}_{o}(s))=\texttt{Dis}_{o}(\mathcal{H}_{o}(t))].\nonumber
\end{align}
\end{definition} 
Intuitively, the above definition says that, 
for any two strings generated by the system, if they cannot be distinguished by the high-level observer, 
then the low-level observer must also have consistent knowledge regarding these two strings. 
Hence, the high-level observer can always determine the knowledge status of the low-level observer. 

\begin{proposition}\upshape
    System $G$ is \emph{high-order detectable} if and only if $G\models\langle  \forall, \texttt{Dis}_o,\texttt{T},\texttt{Y} \rangle$ and $G\models\langle  \forall, \texttt{Dis}_o,\texttt{F},\texttt{N} \rangle$.
\end{proposition}

\begin{pf}
($\Rightarrow$)Note that for any $s\in\mathcal{L}(G)$, either we have $\texttt{Dis}_o(\mathcal{H}_o(s))=\texttt{T}$ or $\texttt{Dis}_o(\mathcal{H}_o(s))=\texttt{F}$ according to the definition of $\texttt{Dis}_o$. Therefore, to satisfy the above requirement, for the case where $\texttt{Dis}_o(\mathcal{H}_o(s))=\texttt{T}$, we have 
\begin{align}
        &(\forall s\in \mathcal{L}(G): \texttt{Dis}_o(\mathcal{H}_o(s))=\texttt{T}) (\forall t\in \mathcal{L}(G))\nonumber\\
        &[\mathcal{H}_{a}(s)=\mathcal{H}_{a}(t)\Rightarrow\texttt{Dis}_{o}(\mathcal{H}_{o}(t))=\texttt{T}],\nonumber
\end{align}
i.e., $G\models\langle  \forall, \texttt{Dis}_o,\texttt{T},\texttt{Y} \rangle$.
Similarly, for the case where $\texttt{Dis}_o(\mathcal{H}_o(s))=\texttt{F}$, we have
\begin{align}
        &(\forall s\in \mathcal{L}(G): \texttt{Dis}_o(\mathcal{H}_o(s))=\texttt{F}) (\forall t\in \mathcal{L}(G))\nonumber\\
        &[\mathcal{H}_{a}(s)=\mathcal{H}_{a}(t)\Rightarrow\texttt{Dis}_{o}(\mathcal{H}_{o}(t))=\texttt{F}],\nonumber
\end{align}
i.e., $G\models\langle  \forall, \texttt{Dis}_o,\texttt{F},\texttt{N} \rangle$.

($\Leftarrow$) We substitute $\langle  \forall, \texttt{Dis}_o,\texttt{T},\texttt{Y} \rangle$ and $\langle  \forall, \texttt{Dis}_o,\texttt{F},\texttt{N} \rangle$ into formula (\ref{eq:epis-prop}). Then we can get the following two formulae.
\begin{align}
        &(\forall s\in \mathcal{L}(G): \texttt{Dis}_o(\mathcal{H}_o(s))=\texttt{T}) (\forall t\in \mathcal{L}(G))\nonumber\\
        &[\mathcal{H}_{a}(s)=\mathcal{H}_{a}(t)\Rightarrow\texttt{Dis}_{o}(\mathcal{H}_{o}(t))=\texttt{T}],\nonumber
\end{align}
and
\begin{align}
        &(\forall s\in \mathcal{L}(G): \texttt{Dis}_o(\mathcal{H}_o(s))=\texttt{F}) (\forall t\in \mathcal{L}(G))\nonumber\\
        &[\mathcal{H}_{a}(s)=\mathcal{H}_{a}(t)\Rightarrow\texttt{Dis}_{o}(\mathcal{H}_{o}(t))=\texttt{F}].\nonumber
\end{align}
Clearly, a system is high-order detectable if these two cases are both satisfied.
\hfill $\qed$    
\end{pf}

\subsection{Other Properties}
Despite the variety of the knowledge $\texttt{Kw}_o$ of the low-level observer by itself, 
in principle, our definition of epistemic properties admits twelve parameter patterns by considering
$\mathbb{Q}\in\{\forall,\exists\}$, $\mathbb{K}_o\in \{\texttt{T},\texttt{F}\}$ 
and 
$\mathbb{K}_{ao}\in \{\texttt{Y},\texttt{N},\texttt{U}\}$. 
However, some patterns are meaningless 
such as $\langle  \mathbb{Q}, \texttt{Kw}_o, \texttt{T}, \texttt{F} \rangle$ or $\langle  \mathbb{Q}, \texttt{Kw}_o, \texttt{N}, \texttt{Y} \rangle$. 
This is because if 
the knowledge of the low-level observer is already true, then the high-level observer is impossible to know for sure that the low-level observer does not have the knowledge. 
Therefore, we will only consider the following eight patterns in the verification problems:  
(1) $\langle  \forall, \texttt{Kw}_o,\texttt{T},\texttt{U} \rangle$;
(2) $\langle  \forall, \texttt{Kw}_o,\texttt{F},\texttt{U} \rangle$;
(3) $\langle  \exists, \texttt{Kw}_o,\texttt{T},\texttt{Y} \rangle$;
(4) $\langle  \exists, \texttt{Kw}_o,\texttt{F},\texttt{N} \rangle$;  
(5) $\langle  \forall, \texttt{Kw}_o,\texttt{T},\texttt{Y} \rangle$;
(6) $\langle  \forall, \texttt{Kw}_o,\texttt{F},\texttt{N} \rangle$;
(7) $\langle  \exists, \texttt{Kw}_o,\texttt{T},\texttt{U} \rangle$; and   
(8) $\langle  \exists, \texttt{Kw}_o,\texttt{F},\texttt{U} \rangle$.

\section{Verification of Epistemic Properties}\label{sec-verify}
In this section, we consider the verification problem of  epistemic properties. 
First,  we provide a general approach using the structure called \emph{double-estimator} to verify all types of epistemic properties. 
This approach is general but has a double-exponential worst-case complexity. 
Then we show that, for some categories of epistemic properties, 
their verification can be done more efficiently with single-exponential. 

\subsection{General Algorithm by Double-Estimators}
Since epistemic properties involve one agent's knowledge of another agent's knowledge, a natural approach is to build two succinct observers (state estimators) to recognize the multi-level inference of states. Here, we demonstrate how this idea can be effectively implemented using the notion of a double-estimator.
 
First, we build a structure called the \emph{knowledge recognizer} to track the value of the predicate $\texttt{Kw}_o$ from the low-level observer's point of view.
\begin{definition}[Knowledge Recognizer]\upshape
    Given system $G$ with observation function $\mathcal{H}_o$, the knowledge recognizer (w.r.t. $\mathcal{H}_o$) is defined as a four-tuple\vspace{-6pt}
\[
T_o=(Q, \Sigma, f, q_0), \vspace{-6pt}
\]
where\vspace{-6pt}  
\begin{itemize}
    \item $Q\subseteq X\times 2^{X}$ is the set of states,\medskip
    \item $\Sigma$ is the set of events,\medskip
    \item $q_0=(x_0,\{\delta(w): w\in \Sigma_{uo}^*\})$ is the initial state, and\medskip
    \item $f: Q\times\Sigma\to Q$ is the deterministic transition function defined by:
    for any $q=(x,y)\in Q$ and $\sigma\in\Sigma$, if $\sigma\in \Sigma_o$, we have\vspace{-6pt}
    \[
    f(q,\sigma)\!=\! 
    \left(\!\!\delta(x,\sigma),\left\{\delta(x,w):
    \begin{array}{cc}
         &  \exists x\in y, w\in \Sigma^*,\\
         &  \mathcal{H}_o(w)=\mathcal{H}_o(\sigma)
    \end{array} \right\}
    \!\!\right);
    \]
    otherwise, we have 
    $ f(q,\sigma)= (\delta(x,\sigma),y)$.
\end{itemize}
\end{definition}

Intuitively, the knowledge recognizer is constructed by synchronizing the original system $G$ with the current-state estimator of itself w.r.t.\ $\mathcal{H}_o$. By construction, for any $s\in\mathcal{L}(G)$, we have $f(s)=(\delta(s),\widehat{X}_o(\mathcal{H}_o(s)))$, where the first element is the current state upon the occurrence of $s$, and the second element is the current-state estimate of $G$ upon observation $\mathcal{H}_o(s)$.
Structure $T_o$ essentially tracks whether the low-level observer knows the fact upon any occurrence of  string $s\in \mathcal{L}(G)$. Therefore,  for any $q=(x,y)\in Q$, we say \vspace{-6pt}
\begin{itemize}
\item 
$q$ is a \emph{known state} iff $\texttt{Kw}_o(y)=\texttt{T}$; and\medskip
\item  
 $q$ is a \emph{unknown state} iff $\texttt{Kw}_o(y)=\texttt{F}$.\vspace{-6pt}
\end{itemize} 
We denote the set of known states  and  unknown states as $Q_{\texttt{T}}\subseteq Q$ and  $Q_{\texttt{F}}:= Q\setminus Q_{\texttt{T}}$, respectively.

For epistemic properties, we need to further think from the high-level observer's point of view, i.e., how the high-level observer estimates the knowledge of the low-level observer. To this end, we build further the estimator automaton of $T_o$ w.r.t. $\Sigma_a$, which is referred to as the \emph{double-estimator}, as follows 

\begin{definition}[Double-Estimator]\upshape
    Given system $G$ with observation functions $\mathcal{H}_o$ and $\mathcal{H}_a$, the double-estimator is defined as a four-tuple\vspace{-6pt}
    \[
        Obs_D(G)=(Q_D, \Delta_a, f_D, q_{0,D}), \vspace{-6pt}
    \]
    where  
    $Q_D\subseteq 2^{Q} \setminus \emptyset$ is the set of states,  
    $\Delta_a$ is the set of observation symbols of the high-level observer,
    $q_{0,D}=\{ f(w)\in Q:   w\in \Sigma_{ua}^*   \}$ is the initial state, 
    and 
    $f_D: Q_D\times \Delta_a\to Q_D$ is the deterministic transition function defined by: 
    for any $q_D\in Q_D$ and $\sigma\in\Sigma_a$,  we have
    \[
    f_D(q_D,\mathcal{H}_a(\sigma))\!=\!
    \{
    f(q, w) \!\in\! Q:   q\!\in\! q_D, \mathcal{H}_a(w)\!=\!\mathcal{H}_a(\sigma)  
    \}.
    \]    
\end{definition}

Intuitively, the double-estimator $Obs_D(G)$  tracks all the possible states in  $T_o$ based on another observation site $\mathcal{H}_a$. In other words, the states of $Obs_D(G)$ are the current-state estimate of knowledge recognizer $T_o$ from the high-level observer's point of view. Specifically, we have the following result.

\begin{proposition}\upshape\label{prop:double-obs}
    Given system $G$ with observation functions $\mathcal{H}_o$ and $\mathcal{H}_a$, for any $\alpha\in \mathcal{H}_a(\mathcal{L}(G))$, we have\vspace{-6pt}
    \begin{itemize}
        \item $\widehat{\texttt{Kw}}_{ao}(\alpha)=\texttt{Y}$ if
        $f_D(\alpha)\subseteq Q_{\texttt{T}}$;\medskip
        \item $\widehat{\texttt{Kw}}_{ao}(\alpha)=\texttt{F}$ if
        $f_D(\alpha)\subseteq Q_{\texttt{F}}$;\medskip
        \item $\widehat{\texttt{Kw}}_{ao}(\alpha)=\texttt{U}$ otherwise.
    \end{itemize}\vspace{-12pt}
\end{proposition} 
\begin{pf}
The proof is provided in the Appendix.
\end{pf}
 
The above proposition immediately indicates that $Obs_D(G)$ allows for easy verification of epistemic properties. Formally, we have the following theorem.

\begin{theorem}\upshape\label{theo:double_obs}
    Given system $G$ with observation functions $\mathcal{H}_o$ and $\mathcal{H}_a$, we have\vspace{-6pt}
    \begin{enumerate}[1)]
        \item  
        $G\models\langle  \forall, \texttt{Kw}_o,\texttt{T},\texttt{U} \rangle$ iff $\forall q_D \in Q_D: q_D\not\subseteq Q_{\texttt{T}}$;\medskip
        \item 
        $G\models\langle  \forall, \texttt{Kw}_o,\texttt{F},\texttt{U} \rangle$ iff $\forall q_D \in Q_D: q_D\not\subseteq Q_{\texttt{F}}$;\medskip
        \item
        $G\models\langle  \exists, \texttt{Kw}_o,\texttt{T},\texttt{Y} \rangle$ iff $\exists q_D \in Q_D: q_D\subseteq Q_{\texttt{T}}$;\medskip
        \item
        $G\models\langle  \exists, \texttt{Kw}_o,\texttt{F},\texttt{N} \rangle$ iff $\exists q_D \in Q_D: q_D\subseteq Q_{\texttt{F}}$;   \medskip
        \item
        $G\models\langle  \forall, \texttt{Kw}_o,\texttt{T},\texttt{Y} \rangle$ iff $\forall q_D \in Q_D: (q_D\subseteq Q_{\texttt{T}})\vee (q_D\cap Q_{\texttt{T}}=\emptyset)$.
        \item
        $G\models\langle  \forall, \texttt{Kw}_o,\texttt{F},\texttt{N} \rangle$ iff $\forall q_D \in Q_D: (q_D\subseteq Q_{\texttt{F}})\vee (q_D\cap Q_{\texttt{F}}=\emptyset)$;\medskip
        \item
        $G\models\langle  \exists, \texttt{Kw}_o,\texttt{T},\texttt{U} \rangle$ iff $\exists q_D \in Q_D: (q_D\not\subseteq Q_{\texttt{T}})\wedge (q_D\cap Q_{\texttt{T}}\ne\emptyset)$; and \medskip     
        \item
        $G\models\langle  \exists, \texttt{Kw}_o,\texttt{F},\texttt{U} \rangle$ iff $\exists q_D \in Q_D: (q_D\not\subseteq Q_{\texttt{F}})\wedge (q_D\cap Q_{\texttt{F}}\ne\emptyset)$. 
    \end{enumerate}\vspace{-12pt}
\end{theorem}
\begin{pf}
The proof is provided in the Appendix.
\end{pf}

In the following example, we illustrate the double-estimator-based verification process by considering the concept of high-order opacity.  

\begin{example}\upshape
\begin{figure}
  \centering
   \begin{tikzpicture}[->,>={Latex}, thick, initial text={}, node distance=1.3cm, initial where=left, thick, base node/.style={circle, draw, minimum size=6mm}]  
   \node[state, initial, base node, ] (0) {$0$};
   \node[state, base node, ] (1) [below of=0] {$1$};
   \node[state, base node, ] (2) [right of=0] {$2$};
   \node[state, base node, ] (3) [below of=2] {$3$};
   \node[state, base node, fill=red] (4) [right of=2] {$4$};
   \node[state, base node, ] (5) [below of=4] {$5$};
   \node[state, base node, ] (6) [right of=4] {$6$};
   \node[state, base node, ] (7) [below of=6] {$7$};
   
   \path[->]
   (0) edge node [yshift=0.2cm] {$c$} (2)
   (2) edge node [yshift=0.2cm] {$b$} (4)
   (4) edge node [yshift=0.2cm] {$d$} (6)
   (0) edge node [xshift=0.2cm] {$a$} (1)
   (1) edge node [yshift=0.2cm] {$b$} (3)
   (3) edge node [yshift=0.2cm] {$b$} (5)
   (5) edge node [yshift=0.2cm] {$a$} (7)
   (7) edge [loop above ] node         [xshift=0.35cm, yshift=-0.1cm]     {$d$} ();
   \draw[->]
   (6) to [bend right=45] node [yshift=0.28cm] {$d$} (4);
   \end{tikzpicture}
   \caption{System $G$ with $\Sigma_o=\{b,d\}$ and  $\Sigma_a=\{a,b\}$.}\label{fig:run-exam-plant} 
\end{figure}
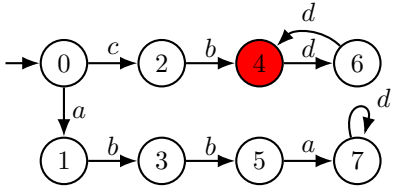
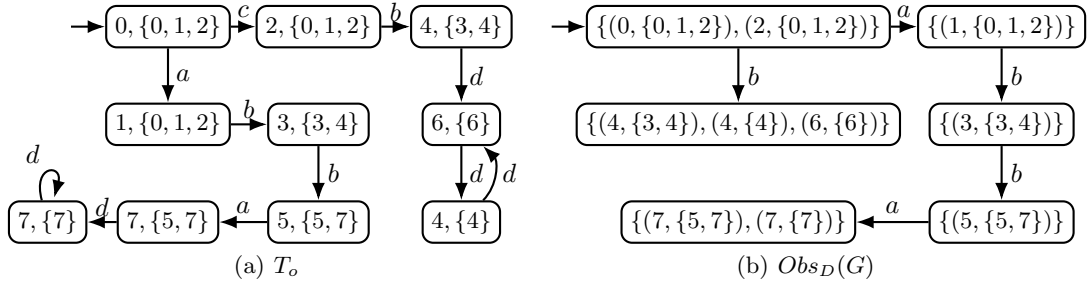
\begin{figure*}[t]
  \centering
    \subfigure[$T_o$\label{fig:know-rec}]{\centering
   \begin{tikzpicture}[->,>={Latex}, thick, initial text={}, node distance=1.3cm, initial where=left, thick, base node/.style={rectangle, rounded corners, draw, minimum size=4mm, font=\footnotesize}]  
   \node[initial, state, base node, rectangle, rounded corners, ] (0) {$0,\{ 0,1,2 \}$} ;
   \node[state, base node, rectangle, rounded corners, ] [xshift=2cm] (1) {$2,\{0,1,2\}$} ;
   \node[state, base node, rectangle, rounded corners, ] [below of=0] (2) {$1,\{0,1,2\}$} ;
   \node[state, base node, rectangle, rounded corners, ] [xshift=3.9cm] (3) {$4,\{3,4\}$} ;
   \node[state, base node, rectangle, rounded corners, ] [below of=1] (4) {$3,\{3,4\}$} ;
   \node[state, base node, rectangle, rounded corners, ] [below of=3] (5) {$6,\{6\}$} ;
   \node[state, base node, rectangle, rounded corners, ] [below of=4] (6) {$5,\{5,7\}$} ;
   \node[state, base node, rectangle, rounded corners, ] [below of=5] (7) {$4,\{4\}$} ;
   \node[state, base node, rectangle, rounded corners, ] [below of=2] (8) {$7,\{5,7\}$} ;
   \node[state, base node, rectangle, rounded corners, ] [xshift=-1.6cm, yshift=-2.6cm] (9) {$7,\{7\}$} ;
   
   \path[->]
   (0) edge node [yshift=0.2cm] {$c$} (1)
   (0) edge node [xshift=0.2cm] {$a$} (2)
   (1) edge node [yshift=0.2cm] {$b$} (3) 
   (2) edge node [yshift=0.2cm] {$b$} (4)
   (3) edge node [xshift=0.2cm] {$d$} (5)
   (5) edge node [xshift=0.2cm] {$d$} (7)
   (4) edge node [xshift=0.2cm] {$b$} (6)
   (6) edge node [yshift=0.2cm] {$a$} (8)
   (8) edge node [yshift=0.2cm] {$d$} (9)
   (9) edge [loop above] node [xshift=-0.2cm]  {$d$} ();
   \draw[->]
   (7) to [bend right=45] node [xshift=0.2cm] {$d$} (5);
   \end{tikzpicture}}
	\subfigure[$Obs_D(G)$	\label{fig:double-obs}]{\centering
   \begin{tikzpicture}[->,>={Latex}, thick, initial text={}, node distance=1.3cm, initial where=left, thick, base node/.style={rectangle, rounded corners, draw, minimum size=4mm, font=\footnotesize}]  
   \node[initial, state, base node, rectangle, rounded corners, ] (0) {$\{(0,\{ 0,1,2 \}),(2,\{ 0,1,2 \})\}$} ;
   \node[state, base node, rectangle, rounded corners, ] [xshift=3.5cm] (1) {$\{(1,\{ 0,1,2 \})\}$} ;
   \node[state, base node, rectangle, rounded corners, ] [below of=0] (2) {$\{(4,\{3,4\}),(4,\{4\}),(6,\{6\})\}$} ;
   \node[state, base node, rectangle, rounded corners, ] [below of=1] (3) {$\{(3,\{ 3,4 \})\}$} ;
   \node[state, base node, rectangle, rounded corners, ] [below of=3] (4) {$\{(5,\{5,7\})\}$} ;
   \node[state, base node, rectangle, rounded corners, ] [below of=2] (5) {$\{(7,\{5,7\}),(7,\{7\})\}$} ;
   \path[->]
   (0) edge node [yshift=0.2cm] {$a$} (1)
   (0) edge node [xshift=0.2cm] {$b$} (2)
   (1) edge node [xshift=0.2cm] {$b$} (3)
   (3) edge node [xshift=0.2cm] {$b$} (4)
   (4) edge node [yshift=0.2cm] {$a$} (5);
   \end{tikzpicture}} \\
  \caption{The construction of $Obs_D(G)$ for $G$.	}
	\label{fig:veri-double-obs}
	\vspace{6pt}
\end{figure*}
Let us consider system $G$ shown in Figure~\ref{fig:run-exam-plant}, 
where the user can observe $\Sigma_o=\{b,d\}$ by $\mathcal{H}_o(b)=b$ and $\mathcal{H}_o(d)=d$, and the intruder can observe $\Sigma_a=\{a,b\}$ by $\mathcal{H}_a(a)=a$ and $\mathcal{H}_a(b)=b$. 
We assume the distinguishment task of the user is to determine the current state of the system, i.e., 
$T_\texttt{spec}=\{(x,x')\in X\times X: x\neq x'\}$. 
Clearly, this system is high-order opaque. 
To see this, we note that the user can determine the current state 
only after it observes the first occurrence of event $d$. 
For string $s=abbad^n$ such that $\texttt{Kw}_o(bbd^n)=\texttt{T}$, the 
intruder observes $H_a(s)=abba$ and it may think that what actually happens is string $s'=abba$
such that $\texttt{Kw}_o(\mathcal{H}_o(abba))=\texttt{F}$ because the user cannot distinguish between states $5$ and $7$ by observing $bb$. 
Similarly, for $t=cbd^{2n}$ such that $\texttt{Kw}_o(bd^{2n})=\texttt{T}$,  the 
intruder observes $\mathcal{H}_a(t)=b$ and it may think that what actually happens is string $t'=cb$
such that $\texttt{Kw}_o(\mathcal{H}_o(cb))=\texttt{F}$ because the user cannot distinguish between states $3$ and $4$ by observing $b$. 
In other words, the intruder can never know that the observer knows the system's current state $G$.

The knowledge recognizer $T_o$ is shown in Figure~\ref{fig:know-rec}. 
Since we consider knowledge task $T_\texttt{spec}=\{(x,x')\in X\times X: x\neq x'\}$, 
for any string $s\in \mathcal{L}(G)$,  we have $\texttt{Kw}_o(\mathcal{H}_o(s))=\texttt{T}$ 
iff the second element of $f(s)$ is a singleton. 
Therefore, we have  $Q_\texttt{T}=\{(4,\{4\}),(6,\{6\}),(7,\{7\})\}$. 
Based on observer $T_o$, we further build the double-estimator $Obs_D(G)$ as shown in Figure~\ref{fig:double-obs}. 
For each state $q\in Q_D$ in it, we see that $q$ always contains an element not in $Q_\texttt{T}$. 
Therefore, we conclude that $G$ is high-order opaque. 
This conclusion is consistent with our above analysis.
\end{example}

Note that the double-estimator structure uses the subset construction technique to capture information uncertainties for both the user and the intruder. Since we use the subset construction twice, the double-estimator is doubly exponential in the size of the plant.

\subsection{Twin-Estimators}
The double-estimator approach described above can be used to verify any epistemic property defined by IS-based knowledge. However, in some scenarios, the high-level observer is primarily interested in determining whether the intruder has sufficient knowledge or not. In such cases, constructing a second estimator for the knowledge recognizer can be unnecessarily costly. Instead, a more efficient approach is to use a \emph{twin-estimator} to capture the desired information.

\begin{definition}[Twin-Estimator]\upshape
Given system $G$ with observation functions $\mathcal{H}_o$ and $\mathcal{H}_a$, the twin-estimator is a four-tuple 
\[
V=(Q_V,\Sigma_V,f_V,q_{0,V}), 
\]
where\vspace{-6pt}
\begin{itemize}
    \item 
    $Q_V\subseteq Q\times Q$ is the set of states;\medskip
    \item 
    $\Sigma_V=\Sigma_V^a\dot{\cup}\Sigma_V^{ua}$ is the set of events, where\medskip
    \begin{itemize}
        \item [-] $\Sigma_V^a=\{(\sigma_1,\sigma_2)\in \Sigma_a\times \Sigma_a: \mathcal{H}_a(\sigma_1)= \mathcal{H}_a(\sigma_2) \}$;\medskip
        \item [-] $\Sigma_V^{ua}=\{(\sigma_1,\epsilon):\sigma_1\in \Sigma_{ua}\}\cup\{(\epsilon,\sigma_2):\sigma_2\in \Sigma_{ua}\}$;\medskip
    \end{itemize}
    \item
    $f_V:Q_V\times \Sigma_V\to Q_V$ is the transition function defined by: for any $q_V=(q_1,q_2)\in Q_V$ and $\sigma_V=(\sigma_1,\sigma_2)\in \Sigma_V$, we have\vspace{-6pt}
    \[
    f_V(q_V,\sigma_V)=(f(q_1,\sigma_1),f(q_2,\sigma_2));\vspace{-6pt}
    \]
    \item 
    $q_{0,V}=(q_0,q_0)$ is the initial state. 
\end{itemize}
\end{definition}

The twin-estimator is constructed by synchronizing the knowledge recognizer $T_o$ with its copy according to the observations under $\mathcal{H}_a$. 
Each state in $V$ is a state pair in $T_o$ and each event in $V$ is also an event pair in $T_o$.  The event set $\Sigma_V$ is divided into $\Sigma_V^a$ and $\Sigma_V^{ua}$, 
and  $(\sigma_1,\sigma_2)$ is in $\Sigma_V^a$ iff $\sigma_1,\sigma_2\in\Sigma_a$ and $\mathcal{H}_a(\sigma_1)=\mathcal{H}_a(\sigma_2)$.
Also, $(\sigma_1,\sigma_2)$ is in $\Sigma_V^{ua}$ iff $\sigma_1(\sigma_2)\in\Sigma_{ua}$ and $\sigma_2(\sigma_1)=\epsilon$. By construction, the twin-estimator has the following two properties. First, for any string $s=(s_1,s_2)\in\mathcal{L}(V)$, we have $\mathcal{H}_a(s_1)=\mathcal{H}_a(s_2)$. Second, for any strings $s_1,s_2\in \mathcal{L}(G)$ such that $\mathcal{H}_a(s_1)=\mathcal{H}_a(s_2)$, there exists a string $s=(s_1,s_2)$ such that $s\in\mathcal{L}(V)$.

For the sake of convenience, for any $q_V=(q_1,q_2)\in Q_V$, we denote $\theta_1(q_V)=q_1$ and $\theta_2(q_V)=q_2$ as the first and second component of $q_V$, respectively. 
Next, we show that, for epistemic properties with the following four-parameter patterns, their verification problems can be solved more efficiently using the twin-estimator.

\begin{theorem}\label{thm:tw}\upshape
    Given system $G$ with observation functions $\mathcal{H}_o$ and $\mathcal{H}_a$, we have 
    \vspace{-6pt}
    \begin{enumerate}[1)]
        \item
        $G\models\langle  \forall, \texttt{Kw}_o,\texttt{T},\texttt{Y} \rangle$ iff for any $q_V\in Q_V$, we have
        $[\theta_1(q_V),\theta_2(q_V)\in Q_{\texttt{T}}]\vee [\theta_1(q_V),\theta_2(q_V)\in Q_{\texttt{F}}]$;\medskip
        \item
        $G\models\langle  \forall, \texttt{Kw}_o,\texttt{F},\texttt{N} \rangle$  iff for any $q_V\in Q_V$, we have
        $ [\theta_1(q_V),\theta_2(q_V)\in Q_{\texttt{T}}]\vee [\theta_1(q_V),\theta_2(q_V)\in Q_{\texttt{F}}]$;\medskip
        \item
        $G\models\langle  \exists, \texttt{Kw}_o,\texttt{T},\texttt{U} \rangle$ iff there exists $q_V\in Q_V$, such that
        $ [\theta_1(q_V)\in Q_{\texttt{T}}]\land [\theta_2(q_V)\in Q_{\texttt{F}}]$; and\medskip
        \item
        $G\models\langle  \exists, \texttt{Kw}_o,\texttt{F},\texttt{U} \rangle$  iff there exists $q_V\in Q_V$, such that
        $[\theta_1(q_V)\in Q_{\texttt{T}}]\land [\theta_2(q_V)\in Q_{\texttt{F}}]$. 
    \end{enumerate}\vspace{-12pt}
\end{theorem}

\begin{pf}
The proof is provided in the Appendix.
\end{pf}

We illustrate the verification procedure using the twin-estimator in the following example.

\begin{example}\upshape
    Let us still consider system $G$ shown in Figure~\ref{fig:run-exam-plant}, here we consider the epistemic diagnosability where we assume $X_S=\{4\}$. The intruder can observe $\Sigma_o=\{b,d\}$, and the user can observe $\Sigma_a=\{a,b\}$. Clearly, this system is not epistemically diagnosable. To see this, let's consider the string $s=cbd^{2n}$ with $n>0$, upon which the intruder observers $bd^{2n}$ and $\hat{X}_o(bd^{2n})=\{4\}$ and $\texttt{Kw}_o(bd^{2n})=\texttt{T}$. However, from the user's point of view, it observes $b$ and may think that $s'=cb$ actually happens. Upon $s'$, we have $\hat{X}_o(b)=\{3,4\}$ and thus $\texttt{Kw}_o(b)=\texttt{F}$. Therefore, the system violates the requirement of epistemic diagnosability upon the occurrence of $s=cbd^{2n}$.

    The knowledge recognizer $T_o$ is shown in Figure~\ref{fig:know-rec}. 
    Since we consider knowledge task $T_\texttt{spec}=(X\setminus X_S)\times(X\setminus X_S)$, for any string $s\in \mathcal{L}(G)$,  we have $\texttt{Kw}_o(\mathcal{H}_o(s))=\texttt{T}$ iff the second element of $f(s)$ is $\{4\}$. Therefore, we have  $Q_\texttt{T}=\{(4,\{4\})\}$. 
    Based on observer $T_o$, we further build the twin-estimator $V$, which is partially shown in Figure~\ref{fig:V}. In this structure, as highlighted by the red color, there is a reachable state $((4,\{ 4 \}),(4,\{ 3,4 \}))$ such that $(4,\{ 4 \})\in Q_\texttt{T}$ and $(4,\{ 3,4 \})\in Q_\texttt{F}$. Therefore, we can conclude that system $G$ is not epistemically diagnosable.
\begin{figure}
    \centering
        \begin{tikzpicture}[->,>={Latex}, thick, initial text={}, node distance=1.3cm, initial where=left, thick, base node/.style={rectangle, rounded corners, draw, minimum size=4mm, font=\footnotesize}]  
        \node[state, initial, base node, ] (0) {$((0,\{ 0,1,2 \}),(0,\{ 0,1,2 \}))$};
        \node[state, base node, ] (1) [below of =0] {$((2,\{ 0,1,2 \}),(0,\{ 0,1,2 \}))$};
        \node[state, base node, ] (2) [below of = 1] {$((2,\{ 0,1,2 \}),(2,\{ 0,1,2 \}))$};
        \node[state, base node, ] (3) [xshift = 4cm, yshift=-2.6cm] {$((4,\{ 3,4 \}),(4,\{ 3,4 \}))$};
        \node[state, base node,] (4) [above of = 3] {$((6,\{ 6 \}),(4,\{ 3,4 \}))$};
        \node[state, base node, red,] (5) [above of = 4] {$((4,\{ 4 \}),(4,\{ 3,4 \}))$};
        
        \draw[->, dashed] (0) -- (-1.5cm, -0.7cm);
        \draw[->, dashed] (4) -- (5.5cm, -0.6cm);
        
        \path[->]
        (0) edge node [xshift=0.4cm] {\footnotesize$(c,\epsilon)$} (1)    (1) edge node [xshift=0.4cm] {\footnotesize$(\epsilon,c)$} (2)
        (2) edge node [yshift=0.4cm] {\footnotesize$(b,b)$} (3)
        (3) edge node [xshift=0.4cm] {\footnotesize$(d,\epsilon)$} (4)
        (4) edge node [xshift=0.4cm] {\footnotesize$(d,\epsilon)$} (5);
        \draw[->]
        (5) to [bend right=45] node [xshift=-0.4cm] {\footnotesize$(d,\epsilon)$} (4);
        \end{tikzpicture}
        \caption{twin-estimator of the system $G$.}
        \label{fig:V}
\end{figure}
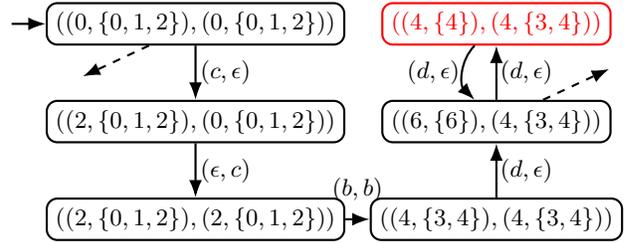
\end{example}

\subsection{State-Pair-Estimators}
Note that in the previous subsection, the twin-estimator explored the structural properties of specific parameter patterns of the high-level observer. In fact, when the parameter patterns of the low-level observer take on certain specific forms, it is possible to further reduce the general verification complexity.

For this purpose, we observe that for the specific kind of IS-based knowledge $\texttt{Dis}_o$, the low-level observer is primarily concerned with whether or not it can distinguish all pairs in $T_\texttt{spec}$. Therefore, for the high-level observer, it is sufficient to estimate the set of all state pairs that the user cannot distinguish. This leads to the concept of the state-pair-estimator, which also operates with only a single exponential complexity.

\begin{definition}[State-Pair-Estimator]\upshape
    Given system $G$ with observation functions $\mathcal{H}_o$ and $\mathcal{H}_a$, the state-pair-estimator is defined as a four-tuple\vspace{-6pt}
    \[
    Obs_P(G)=( Q_P, \Delta_a,  f_P, q_{0,P}), \vspace{-6pt}
    \] 
where\vspace{-6pt}
\begin{itemize}
    \item 
    $Q_P \subseteq 2^{X\times X^2}\setminus\emptyset$ is the set of states;\medskip
    \item 
    $\Delta_a$ is the set of intruder's observation symbols;  \medskip
    \item 
    $f_P: Q_P\times \Delta_a \to Q_P$ is the deterministic transition function defined by: 
    for any $q_P\in Q_P, \sigma\in \Sigma_a$, we have\vspace{-6pt}
    \begin{align}
     &f_P(q_P,\mathcal{H}_a(\sigma))=     \nonumber\\
    & \left\{
     (x',(x_1',x_2')) :\!
     \begin{array}{cc}
       \exists (x,(x_1,x_2))\in q_P, \\ 
       \exists w, w_1,w_2\in  \Sigma^* \text{ s.t.} \\
       \mathcal{H}_a(w)=\mathcal{H}_a(\sigma), \\
       \mathcal{H}_o(w)=\mathcal{H}_o(w_1)=\mathcal{H}_o(w_2)\\
       \text{and } 
       (x',(x_1',x_2'))=\\
       (\delta(x,w),(\delta(x_1,w_1),\delta(x_2,w_2)))
    \end{array} 
     \right\}, 
    \nonumber
    \end{align}
    \item 
    $q_{0,P}$ is the unique initial state defined by:\vspace{-6pt}
     \begin{align}
     &q_{0,P}=     \nonumber\\
    & \left\{\!\!
     (x',(x_1',x_2')) :\!\!
     \begin{array}{cc}
       \exists w\! \in\!  \Sigma_{ua}^*,  
       w_1,w_2\! \in\!   \Sigma^* \text{ s.t. } \\
       \mathcal{H}_o(w)=\mathcal{H}_o(w_1)=\mathcal{H}_o(w_2)\\
       \text{and }
       (x',(x_1',x_2'))=\\
       (\delta(w),(\delta(w_1),\delta(w_2)))
    \end{array} \! \! 
     \right\}.
    \nonumber
    \end{align}
\end{itemize}
\end{definition}

Intuitively, the state-pair-estimator estimates both the current state of the system and the set of state pairs that the low-level observer cannot distinguish from the high-level observer's point of view. 
Specifically, suppose that the current estimate of the high-level observer is $q_P\in 2^{X\times X^2}$, the meaning of $(x,(x_1,x_2))\in q_P$ is that the high-level observer thinks that the system is currently in $x$ and meanwhile, it thinks that the low-level observer cannot distinguish state pair $(x_1,x_2)$. 
By observing a new event $\sigma \in \Sigma_a$, the high-level observer updates its estimate by considering all strings $w\in \Sigma^*$ such that $\mathcal{H}_a(w)=\mathcal{H}_a(\sigma)$, and $\delta(x,w)!$. 
Note that, for each actual string $w$ in the system, the low-level observer has its observation through $\mathcal{H}_o$. 
Then for strings $w_1\in \mathcal{L}(G,x_1)$ and $w_2\in \mathcal{L}(G,x_2)$, if $\mathcal{H}_o(w_1)=\mathcal{H}_o(w_2)=\mathcal{H}_o(w)$,
then the intruder thinks that the user again cannot distinguish between state $\delta(x_1,w_1)$ and state $\delta(x_2,w_2)$.

For any state $q_P\in 2^{X\times X^2}$, we say $q_P$ is a \emph{high-level known state} if 
for any state $(x,(x_1,x_2))\in q_P$, we have $(x_1,x_2)\notin T_\texttt{spec}$. We denote the set of high-level known states as\vspace{-6pt}
\[
Q^{\texttt{Y}}_P= \{ q_P\in Q_P:    q_P\cap (X\times T_\texttt{spec}) =\emptyset \}.\vspace{-6pt}
\]
Now we show that some classes of epistemic properties can be efficiently checked by the state-pair-estimator.  

\begin{theorem}\label{thm:spair}\upshape
    Given system $G$ with observation functions $\mathcal{H}_o$ and $\mathcal{H}_a$, we have 
    \vspace{-6pt}
    \begin{itemize}
        \item[1)] $G\models\langle  \forall, \texttt{Dis}_o,\texttt{T},\texttt{U} \rangle$ iff $Q^{\texttt{Y}}_P=\emptyset$; and\medskip
        \item[2)] $G\models\langle  \exists, \texttt{Dis}_o,\texttt{T},\texttt{Y} \rangle$ iff $Q^{\texttt{Y}}_P\ne\emptyset$.
    \end{itemize}\vspace{-12pt}
\end{theorem}

\begin{pf}
The proof is provided in the Appendix.
\end{pf}

We illustrate the verification procedure using the state-pair-estimator in the following example.
\begin{example}\upshape
\begin{figure}
  \centering
	\begin{tikzpicture}[->,>={Latex}, thick, initial text={}, node distance=1.3cm, initial where=left, thick, base node/.style={rectangle, rounded corners, draw, minimum size=4mm, font=\footnotesize}]  
   \node[initial, state, base node, rectangle, rounded corners, ] (0) {$\{0,2\}\times A$} ;
   \node[state, base node, rectangle, rounded corners, ] [xshift=2cm] (1) { $\{1\}\times A$ } ;
   \node[state, base node, rectangle, rounded corners, ] [xshift=4cm] (2) { $\{3\}\times B$ } ;
   \node[state, base node, rectangle, rounded corners, ] [xshift=-1cm, yshift=-1.3cm] (3) { $ \{(6,(6,6))\}\cup(\{4\}\times B)$ } ;
   \node[state, base node, rectangle, rounded corners, ] [below of=2] (4) { $\{5\}\times C$ } ;
   \node[state, base node, rectangle, rounded corners, ] [below of=1] (5) { $\{7\}\times C$ } ;
   \path[->]
   (0) edge node [yshift=0.2cm] {$a$} (1)
   (1) edge node [yshift=0.2cm] {$b$} (2)
   (0) edge node [xshift=0.2cm] {$b$} (3)
   (2) edge node [xshift=0.2cm] {$b$} (4)
   (4) edge node [yshift=0.2cm] {$a$} (5);
   \end{tikzpicture}
  \caption{State-pair-estimator $Obs_P(G)$ for $G$.}
	\label{fig:sp-obs}
	\vspace{6pt}
\end{figure}
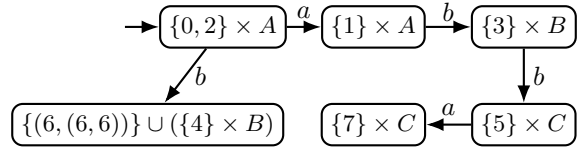
    Let us still consider system $G$ shown in Figure~\ref{fig:run-exam-plant}, consider high-order opacity with $T_\texttt{spec}=\{(x,x')\in X\times X: x\neq x'\}$. 
    This state-pair-estimator is shown in Figure~\ref{fig:sp-obs}, where $A=\{0,1,2\}^2$, $B=\{3,4\}^2$, and $C=\{5,7\}^2$. 
    Initially, we have $\mathcal{H}_a(\epsilon)=\mathcal{H}_a(c)=\epsilon$. 
    Note that we also have $\mathcal{H}_o(\epsilon)=\mathcal{H}_o(c)=\epsilon$
    and $\mathcal{H}_o^{-1}(\epsilon )\cap \mathcal{L}(G)=\{\epsilon, a,c\}$, 
    i.e., the intruder thinks that the user cannot distinguish 
    states $\delta(\epsilon)=0,\delta(a)=1$ and $\delta(c)=2$. 
    Therefore, the $2\times 3\times 3$ combinations of the state pairs give the initial state in $Obs_P(G)$. 
    Also, for example, consider string $s=cbdd$, where  $\mathcal{H}_a(s)=b$. 
    We have $\mathcal{H}_a^{-1}(\mathcal{H}_a(s))\cap \mathcal{L}(G)=\{  cbd^n \}$. 
    For strings  $cbd^n$ where $n\geq 1$, we have $\mathcal{H}_o^{-1}(\mathcal{H}_o(cbd^n))\cap \mathcal{L}(G)= \{cbd^n\}$, i.e., the user can perfectly determine the state, and  corresponding state pairs the user cannot distinguish are $(4,4)$ and $(6,6)$. 
    For strings $cb$, we have $\mathcal{H}_o^{-1}(\mathcal{H}_o(cb))\cap \mathcal{L}(G)= \{ab, cb\}$. Therefore, the state pairs the user cannot distinguish are $(3,3),(4,4),(3,4)$, and $(4,3)$.
    This is why we have $f_P(\mathcal{H}_a(s))=\{(6,(6,6))\}\cup(\{4\}\times B)$.
    Note that, for each state $q_P\in Q_P$, $q_P$ contains a state pair $(x,(x_1,x_2))$ such that $x_1\neq x_2$, which means that 
    $q_P\cap X\times T_\texttt{spec}\neq \emptyset$. Intuitively, this means that the intruder thinks that the user may not be able to distinguish between different states $x_1$ and $x_2$. 
    Since  $Q_P^y=\emptyset$, we conclude that the system $G$ is high-order opaque, which is consistent with our previous analysis.
\end{example}

\section{Epistemic Properties with Delayed Information}\label{sec-delay}
In our definition of epistemic properties, the knowledge of the high-level observer is considered in an ``instant fashion".
For instance, in epistemic diagnosability, it is required that the high-level observer can immediately detect when the low-level observer has released its secret. However, this requirement can be further relaxed by allowing the high-level observer to acquire sufficient knowledge within a finite delay after the low-level observer gains some knowledge.

Here, instead of presenting a general definition of epistemic properties with delayed information, we will use epistemic diagnosability as a specific example to illustrate how this generalization can be achieved.
To this end, we define $\Psi(\texttt{Rev}_o)$ as the set of all finite strings where the critical information leakage occurs for the first time, i.e.,
\[
\Psi(\texttt{Rev}_o):=
\left\{s\sigma\in \mathcal{L}(G)\!\!:\!
\begin{array}{cc}
     & \sigma\in \Sigma_a \land \\
     &\texttt{Rev}_o(\mathcal{H}_o(s\sigma))=\texttt{T} \land \\
     & \texttt{Rev}_o(\mathcal{H}_o(s))=\texttt{F}
\end{array} 
\right\},
\]
Then we can describe the relaxed diagnosis requirement with delayed information as follows.

\begin{definition}[Finite Epistemic Diagnosability]\upshape
Given system $G$, a set of secret states $X_S \subseteq X$ and observation functions $\mathcal{H}_o$ and $\mathcal{H}_a$, system $G$ is said to be \emph{finitely epistemic diagnosable} (w.r.t.\ $X_S$, $\Sigma_a$ and $\Sigma_o$) if 
\begin{align}
    &(\exists n\!\in\!\mathbb{N})(\forall s\!\in\! \Psi(\texttt{Rev}_o)) (\forall t\!\in \!\mathcal{L}(G)/s: |\mathcal{H}_a(t)|\geq n) \nonumber\\
    &(\forall w\!\in\! \mathcal{L}(G))
    [\mathcal{H}_a(st)\!=\!\mathcal{H}_a(w)\Rightarrow \texttt{Rev}_o(\mathcal{H}_o(w))\!=\!\texttt{T} ]. \nonumber
\end{align}
\end{definition}

The verification of finite epistemic diagnosability is similar to its instant counterpart, utilizing the twin-estimator structure. 
However, instead of merely checking for the existence of particular states in the twin-estimator, here we need to search  for the existence of specific cycles, as delayed information is permitted. This change accounts for the time delays allowed for the high-level observer to gain sufficient knowledge after the low-level observer has obtained some information.
 
\begin{theorem}\upshape
        System $G$ is not \emph{finitely epistemically diagnosable}  (w.r.t.\ $X_S$ $\Sigma_a$, and $\Sigma_o$) iff in the twin-estimator, there exists a reachable cycle\vspace{-6pt}
        \[
        q_{1,V}\overset{\sigma_V^1}{\longrightarrow}q_{2,V}\overset{\sigma_V^2}{\longrightarrow}\cdots\overset{\sigma_V^{n-1}}{\longrightarrow}q_{n,V}\vspace{-6pt}
        \]
        such that
        \begin{enumerate}[1)]
            \item
            $\theta_1(q_{i,V})\in Q_{\texttt{T}}$ and $\theta_2(q_{i,V})\in Q_{\texttt{F}}$ for all $i=0,1,...,n$; and\medskip
            \item
            $\sigma_V^j\in \Sigma_V^o$ for some $j=0,1,...,n$.
        \end{enumerate}\vspace{-12pt}
\end{theorem}

\begin{pf}
    ($\Rightarrow$)
    Assume there exists a reachable cycle $\pi$ satisfying 1) and 2), but the system is epistemically diagnosable. 
    Since $\pi$ is a reachable cycle, we can find a string $s_{1,V}\in\mathcal{L}(V)$ such that $q_{1,V}=f_V(s_{1,V})$.
    After repeating the cycle an infinite number of times, we obtain an infinite string $s_V=s_{1, V}(s_{2, V})^{\omega}$.
    Without loss of generality, let $s_V=(s_1,s_2)$, and $\sigma_V^i=(\sigma_1^i,\sigma_2^i)$, $q_{i,V}=(q_{i,1},q_{i,2})$ for every $i$. 
    By conditions 1), for some prefix $t_1\in\overline{\{s_1\}}$, we have $f(t_1)=q_{i,1}\in Q_{\texttt{T}}$. This means that $\texttt{Rev}_o(t_1)=\texttt{T}$ and $\texttt{Rev}_o(s_1)=\texttt{T}$. Also, we know that for any prefix $t_2\in \overline{\{s_2\}}$, we have $f(t_2)\not\in Q_{\texttt{T}}$, which means $\texttt{Rev}_o(s_2)=\texttt{F}$. By condition 2), we know that $|\mathcal{H}_a(s_1)|=\infty$. By the structure of $V$, we have $\mathcal{H}_a(s_1)=\mathcal{H}_a(s_2)$. This is a contradiction of the fact that the system is finitely epistemically diagnosable.
    
    ($\Leftarrow$)Assume $G$ is not finitely epistemically diagnosable. By definition, we know that for any $n\in\mathbb{N}$, there exists a string $t_1\in\Psi(\texttt{Rev}_a)$ and a suffix $w_1$, such that $|\mathcal{H}_a(w_1)|>n$ and $\texttt{Rev}_o(t_1w_1)=\texttt{T}$. Then we have $f(t_1)\in Q_{\texttt{T}}$, $f(t_1 w_1)\in Q_{\texttt{T}}$ and $f(t)\not\in Q_{\texttt{T}}$ for any $t\in\overline{\{t_1\}}\setminus\{t_1\}$. 
    Furthermore, there exists a string $s_2$ with $\mathcal{H}_a(s_2)=\mathcal{H}_a(t_1w_1)$, such that $\texttt{Rev}_o(s_2)=\texttt{F}$, i.e., $f(s_2)\not\in Q_{\texttt{T}}$. Then we have that for all the prefix $t$ of $s_2$, $f(t)\not\in Q_{\texttt{T}}$. Finally, by the structure of $V$, we know that with string $s_V=(t_1w_1,s_2)$, we can get $q_{i,1}\in Q_t$ for all $i>|t_1|$ and $q_{j,2}\not\in Q_{\texttt{T}}$ for all $j\leq |s_2|$. Since $n$ can be arbitrarily large and $V$ is a finite state automaton, by the Pigeonhole principle, we can always find a reachable cycle such that 1) and 2) are both satisfied.
\hfill $\qed$
\end{pf}

\section{Conclusion}\label{sec-con}
In this paper, we introduced a framework for investigating information-flow properties in partially-observed discrete event systems  from a new perspective. Specifically, we focused on a two-observer setting, where the high-level observer aims to infer the knowledge of the low-level observer. We provided a generic definition of epistemic properties, parameterized by quantifiers and knowledge requirements. We show this definition can capture a wide range of useful requirements that naturally arise in scenarios involving multiple observation sites.
There are many future directions for the proposed framework, and we highlight two immediate avenues for exploration. First, it is worthwhile to generalize the framework to accommodate an arbitrary number of observers, enabling the analysis of information structures such as ``agent 1 knows that agent 2 knows that agent 3 does not know". Second, we would like to investigate the control synthesis problem in addition to the verification problem. This would involve synthesizing feedback control laws to regulate the system behavior in such a way that the desired epistemic properties are satisfied.


\section{Appendix}

\textbf{Proof of Proposition~\ref{prop:double-obs}}\vspace{-12pt}
\begin{pf}
To prove the above proposition, we first need to prove that $\texttt{Kw}_o(\mathcal{H}_o(s))=\texttt{T}$ is equivalent to $f(s)\in Q_{\texttt{T}}$, i.e., $\widehat{X}_{o}(\mathcal{H}_o(s))\subseteq X^K$ is equivalent to $f(s)\in Q_{\texttt{T}}$. More specifically, we need to prove that for $f(s)=(x,y)$, we have $y=\widehat{X}_{o}(\mathcal{H}_o(s))$, which we have stressed above. Here we prove by induction on the length of $\mathcal{H}_o(s)$.

\emph{Induction Basis}:
Suppose that $|\mathcal{H}_o(s)|=0$, i.e., $\mathcal{H}_o(s)=\epsilon$. Then we have $y=\{\delta(w):w\in\Sigma^*_{uo}\}=\widehat{X}_{o}(\epsilon)$. It is clear that the induction basis holds.

\emph{Induction Step}:
Now we assume that $y=\widehat{X}_{o}(\mathcal{H}_o(s))$ holds for $|\mathcal{H}_o(s)|=k$. Then we discuss the case of $s\sigma$, where $\sigma\in \Sigma_o$. By the transition function, we have that
\begin{align}
    y'
    & =\{\delta(x,w):\exists x\in y,w\in\Sigma^*,\mathcal{H}_o(w)=\mathcal{H}_o(\sigma)\} \nonumber \\
    & =\left\{\delta(x,w):
    \begin{array}{cc}
         &  \exists x\in \widehat{X}_{o}(\mathcal{H}_o(s)), w\in \Sigma^*,\\
         &  \mathcal{H}_o(w)=\mathcal{H}_o(\sigma)
    \end{array} \right\} \nonumber \\
    & =\{\delta(w')\in X: w'\in \Sigma^*, \mathcal{H}_o(w')=\mathcal{H}_o(s)\mathcal{H}_o(\sigma)\} \nonumber \\
    & = \widehat{X}_{o}(\mathcal{H}_o(s\sigma)) \nonumber
\end{align}
Thus we can conclude that $y=\widehat{X}_{o}(\mathcal{H}_o(s))$ also holds for $|\mathcal{H}_o(s)|=k+1$.
This completes the induction step. 

Now we are ready to prove our result.  According to our proof above and the definition of $\widehat{\texttt{Kw}}_{ao}(\alpha)$, we only need to prove that $f_D(\alpha)= \{f(s)\in D: \mathcal{H}_a(s)=\alpha\}$.
Also, we prove by induction on the length of $\alpha$.

\emph{Induction Basis}:
Suppose that $|\alpha|=0$, i.e., $\alpha=\epsilon$. Then we have $f_D(\alpha)=q_{0,D}=\{ f(w)\in Q:   w\in \Sigma_{ua}^*   \}$. It is clear that the induction basis holds.

\emph{Induction Step}:
Now we assume that the conclusion holds for $|\alpha|=k$. Then we discuss the case of $\alpha\mathcal{H}_a(\sigma)$, where $\sigma\in \Sigma_o$. By the transition function, we have that
\begin{align}
    f_D(& f_D(\alpha),\mathcal{H}_a(\sigma)) \nonumber \\
    & =\{f(q, w) \in Q:  q\in f_D(\alpha), \mathcal{H}_a(w)=\mathcal{H}_a(\sigma)\} \nonumber \\
    & =\{f(w')\in Q: w'\in \Sigma^*, \mathcal{H}_a(w')=\alpha\mathcal{H}_a(\sigma)\} \nonumber 
\end{align}
Thus we can conclude that $f_D(\alpha)= \{f(s)\in D: \mathcal{H}_a(s)=\alpha\}$ also holds for $|\alpha|=k+1$.
This completes the induction step. 
\hfill $\qed$
\end{pf}

\textbf{Proof of Theorem~\ref{theo:double_obs}} \vspace{-12pt}
\begin{pf}
To prove Theorem~\ref{theo:double_obs}, we first note that 1), 2) can be proved equivalently since the definition of the knowledge of the low-level observer $\texttt{T}$ and $\texttt{F}$ are symmetric. Moreover, 3) and 4) are contrapositive propositions of 1) and 2), respectively. Therefore, to prove 1), 2), 3), and 4), we only need to prove 1). It is also the case for 5), 6), 7), and 8). Therefore, we only need to prove 1) and 5).

\emph{Proof of 1)}:
($\Rightarrow$)
By contraposition.  
Suppose a state $q\in Q_D$ exists, such that $q\subseteq Q_{\texttt{T}}$. Then for any $s\in\mathcal{L}(G)$ such that  $f_D(\mathcal{H}_a(s))=q$ we have $f(s)\in Q_{\texttt{T}}$, which means that $\texttt{Kw}_o(\mathcal{H}_o(s))=\texttt{T}$. Since $q\subseteq Q_{\texttt{T}}$, we know that there exists no string $t\in\mathcal{L}(G)$ such that $f(t)\notin Q_{\texttt{T}}$ and $\mathcal{H}_a(t)=\mathcal{H}_a(s)$. In other words, for any $t\in \mathcal{L}(G)$ such that $\mathcal{H}_a(s)= \mathcal{H}_a(t)$, we have $\texttt{Kw}_o(\mathcal{H}_o(t))=\texttt{T}$, which means the system does not satisfy property $\langle  \forall, \texttt{Kw}_o,\texttt{T},\texttt{U} \rangle$.
    
($\Leftarrow$) 
Also by contraposition. 
Suppose that the system does not satisfy $\langle  \forall, \texttt{Kw}_o,\texttt{T},\texttt{U} \rangle$, which means that there exists a string $s\in\mathcal{L}(G)$ such that $\texttt{Kw}_o(\mathcal{H}_o(s))=\texttt{T}$ and $\widehat{\texttt{Kw}}_{ao}(\mathcal{H}_a(s))=\texttt{Y}$. 
Therefore, we have $f(s)\in Q_{\texttt{T}}$ and there exists no string $t\in\mathcal{L}(G)$ such that $\texttt{Kw}_o(\mathcal{H}_o(t))=\texttt{F}$ and $\mathcal{H}_a(t)=\mathcal{H}_a(s)$. This means that for any $t\in\mathcal{L}(G)$ satisfies $\mathcal{H}_a(t)=\mathcal{H}_a(s)$, we have $f(t)\in Q_{\texttt{T}}$. Therefore, we have $f_D(\mathcal{H}_a(s))\subseteq Q_{\texttt{T}}$.

\emph{Proof of 5)}:
($\Rightarrow$)
Assume that there exists a state $q_D\in Q_D$ satisfying that $q_D\cap Q_{\texttt{T}}\ne \emptyset$ and $q_D\not\subseteq Q_{\texttt{T}}$, but the system still satisfying $\langle  \forall, \texttt{Kw}_o,\texttt{T},\texttt{Y} \rangle$.
Then we can find a string $s\in\mathcal{L}(G)$ such that $q_{D}=f_D(\mathcal{H}_a(s))$.
Over $\mathcal{H}_a(s)$, according to Proposition~\ref{prop:double-obs}, we have $\widehat{\texttt{Kw}}_{ao}(\mathcal{H}_a(s))=\texttt{U}$. This means that we can always find $t\in \mathcal{L}(G)$ with $\mathcal{H}_a(s)=\mathcal{H}_a(t)$ satisfying $\texttt{Kw}_o(\mathcal{H}_o(t))=\texttt{F}$.
This is a contradiction of the fact that the system satisfies $\langle  \forall, \texttt{Kw}_o,\texttt{T},\texttt{Y} \rangle$.

    ($\Leftarrow$)Assume $G$ does not satisfy $\langle  \forall, \texttt{Kw}_o,\texttt{T},\texttt{Y} \rangle$. By definition, we know there exists a string $s\in\mathcal{L}(G)$, such that $\texttt{Kw}_o(\mathcal{H}_o(s))=\texttt{T}$. According to Proposition~\ref{prop:double-obs}, we have $f(s)\in Q_{\texttt{T}}$. Furthermore, we know there exists a string $t$ with $\mathcal{H}_a(s)=\mathcal{H}_a(t)$, such that $\texttt{Kw}_o(\mathcal{H}_o(t))=\texttt{F}$, i.e., $f(t)\not\in Q_{\texttt{T}}$. By the structure of the double observer, we know that     $f_D(\mathcal{H}_a(s))=f_D(\mathcal{H}_a(t))\not\subseteq Q_{\texttt{T}}$, also $f_D(\mathcal{H}_a(s))\cap Q_{\texttt{T}}\ne \emptyset$. 
\hfill $\qed$
\end{pf}

\textbf{Proof of Theorem~\ref{thm:tw}}\vspace{-12pt}

\begin{pf}
Since 1) and 2) are equivalent to each other, while 3) and 4) are contrapositive with 1) and 2) respectively, we only need to prove 1).

($\Rightarrow$)
Assume there exists a state $q_V$ such that $(\theta_1(q_V)\in Q_{\texttt{T}})\land (\theta_2(q_V)\in Q_{\texttt{F}})$, but the system also satisfies $\langle  \forall, \texttt{Kw}_o,\texttt{T},\texttt{Y} \rangle$. 
Then we can find a string $s_{V}\in\mathcal{L}(V)$ such that $q_{V}=f_V(s_{V})$.
We denote $s_V=(s_1,s_2)$, then we have $f(s_1)\in Q_{\texttt{T}}$. This means that $\texttt{Kw}_o(s_1)=\texttt{T}$. Also, we know that $f(s_2)\in Q_{\texttt{F}}$, which means $\texttt{Kw}_o(s_2)=\texttt{F}$. 
By the structure of $V$, we have $\mathcal{H}_a(s_1)=\mathcal{H}_a(s_2)$. This   contradicts  fact that the system satisfies $\langle  \forall, \texttt{Kw}_o,\texttt{T},\texttt{Y} \rangle$.
    
($\Leftarrow$)Assume $G$ does not satisfy $\langle  \forall, \texttt{Kw}_o,\texttt{T},\texttt{Y} \rangle$. By definition, we know that there exists a string $s_1\in\mathcal{L}(G)$, such that $\texttt{Kw}_o(s_1)=\texttt{T}$. Then we have $f(s_1)\in Q_{\texttt{T}}$. 
Furthermore, there exists a string $s_2$ with $\mathcal{H}_a(s_2)=\mathcal{H}_a(s_1)$, such that $\texttt{Kw}_o(s_2)=\texttt{F}$, i.e., $f(s_2)\in Q_{\texttt{F}}$. Finally, by the structure of $V$, we know that with string $s_V=(s_1,s_2)$, we can get $\theta_1(f_V(s_V))\in Q_{\texttt{T}}$ and $\theta_1(f_V(s_V))\in Q_{\texttt{F}}$. Which is a contradiction of the fact that $\forall q_V\in Q_V: (\theta_1(q_V),\theta_2(q_V)\in Q_{\texttt{T}})\vee (\theta_1(q_V),\theta_2(q_V)\in Q_{\texttt{F}})$.
\hfill $\qed$
\end{pf}

\textbf{Proof of Theorem~\ref{thm:spair}} \vspace{-12pt}

\begin{pf}
    Note that we only need to prove 1) because 1) and 2) are contrapositive with each other.
    We first prove that for any $s\in \mathcal{L}(G)$, the state reached by $\mathcal{H}_a(s)$ in $Obs_P(G)$ satisfies the following by induction on $|\mathcal{H}_a(s)|$:
    \begin{align}
       & f_P(q_{0,P},\mathcal{H}_a(s))=    \nonumber  \\
       & \left\{
       (\delta(t),(\delta(w_1),\delta(w_2)))  :
        \begin{array}{cc}
           \exists  t, w_1, w_2\in  \mathcal{L}(G), \\
           \text{s.t. } \mathcal{H}_a(s)=\mathcal{H}_a(t)\text{ and}\\ 
               \mathcal{H}_o(t)=\mathcal{H}_o(w_1)=\mathcal{H}_o(w_2) 
         \end{array} 
             \right\}
        \nonumber\vspace{-60pt}
    \end{align} 

    \emph{Induction Basis: } 
    When $\mathcal{H}_a(s)=\epsilon$, we have $f_P(q_{0,P},\epsilon)= q_{0,P}$; clearly, the induction basis holds. 

    \emph{Induction Step}:
    Now we assume that our claim holds for $|\mathcal{H}_a(s)|=k$. Then we discuss the case of $s\sigma$, where $\sigma\in \Sigma_a$. By the transition function, we have that
    \begin{align}
        &f_P(f_P(\mathcal{H}_a(s)),\mathcal{H}_a(\sigma))     \nonumber\\
        &=\left\{
         (x',(x_1',x_2')) :\!
         \begin{array}{cc}
           \exists (x,(x_1,x_2))\in f_P(\mathcal{H}_a(s)), \\ 
           \exists w, w_1,w_2\in  \Sigma^* \text{ s.t.} \\
           \mathcal{H}_a(w)=\mathcal{H}_a(\sigma), \\
           \mathcal{H}_o(w)=\mathcal{H}_o(w_1)=\mathcal{H}_o(w_2)\\
           \text{and } 
           (x',(x_1',x_2'))=\\
           (\delta(x,w),(\delta(x_1,w_1),\delta(x_2,w_2)))
        \end{array} 
         \right\} \nonumber
    \end{align}
    \begin{align}
        &=\left\{
         (x',(x_1',x_2')) :\!\!\!\!\!\!\!\!
         \begin{array}{cc}
           \exists t,w_1,w_2\in \mathcal{L}(G), \\ 
           \exists w, w'_1,w'_2\in  \Sigma^* \text{ s.t.} \\
           \mathcal{H}_a(w)=\mathcal{H}_a(\sigma), \mathcal{H}_a(s)=\mathcal{H}_a(t),\\
           \mathcal{H}_o(w)=\mathcal{H}_o(w'_1)=\mathcal{H}_o(w'_2), \\
           \mathcal{H}_o(t)=\mathcal{H}_o(w_1)=\mathcal{H}_o(w_2), \\
           \text{and } 
           (x',(x_1',x_2'))=\\
           (\delta(tw),(\delta(w_1 w'_1),\delta(w_2 w'_2)))
        \end{array} 
         \right\} \nonumber \\
        &=\left\{
         (x',(x_1',x_2')) :\!\!\!
         \begin{array}{cc}
           \exists tw,w_1 w'_1,w_2 w'_2\in \mathcal{L}(G) \text{ s.t.} \\
           \mathcal{H}_a(tw)=\mathcal{H}_a(s\sigma), \\
           \mathcal{H}_o(tw)=\mathcal{H}_o(w_1 w'_1)=\mathcal{H}_o(w_2 w'_2) \\
           \text{and }
           (x',(x_1',x_2'))=\\
           (\delta(tw),(\delta(w_1 w'_1),\delta(w_2 w'_2)))
        \end{array} 
         \right\} \nonumber 
    \end{align}
Thus we can conclude that our claim above also holds for $|\mathcal{H}_a(s)|=k+1$.
This completes the induction step.     

Now we prove the theorem by contraposition.

($\Rightarrow$)
Suppose that $Q_P^y\ne\emptyset$, i.e., there exists a state $q_P\in Q_P$ such that $q_P\cap (X\times T_\texttt{spec}) =\emptyset$.
Let us consider the string $s\in\mathcal{L}(G)$ such that  $f_P(q_{0,P},\mathcal{H}_a(s))=q_P$. Then for any $t, w_1, w_2\in \mathcal{L}(G)$ such that $\mathcal{H}_o(t)=\mathcal{H}_o(w_1)=\mathcal{H}_o(w_2)$ and $\mathcal{H}_a(s)=\mathcal{H}_a(t)$, we have $(\delta(t),(\delta(w_1),\delta(w_2)))\notin X\times T_\texttt{spec}$, i.e., $(\delta(w_1),\delta(w_2))\notin T_\texttt{spec}$. Furthermore, the product of the state estimation of $\mathcal{H}_o(t)$ is
    \begin{align}
     &\hat{X}_o(\mathcal{H}_o(t))\times \hat{X}_o(\mathcal{H}_o(t))=     \nonumber\\
    & \left\{ 
     (\delta(w_1),\delta(w_2)) : 
     \begin{array}{cc}
       \exists w_1,w_2\! \in\!   \mathcal{L}(G) \text{ s.t.} \\
       \mathcal{H}_o(t)=\mathcal{H}_o(w_1)=\mathcal{H}_o(w_2)
    \end{array}  
     \right\}.
    \nonumber
    \end{align}
Therefore, $\forall t\in \mathcal{L}(G):\mathcal{H}_a(t)=\mathcal{H}_a(s)$, $\texttt{Dis}_o(\mathcal{H}_o(t))=\texttt{T}$. This means $G$ does not satisfy $\langle  \forall, \texttt{Dis}_o,\texttt{T},\texttt{U} \rangle$.

($\Leftarrow$)  
Suppose that the system does not satisfy $\langle  \forall, \texttt{Dis}_o,\texttt{T},\texttt{U} \rangle$. 
Then there exists a string $s\in\mathcal{L}(G)$ such that 
\[ 
\forall t\in\mathcal{L}(G): \mathcal{H}_a(t)=\mathcal{H}_a(s)\Rightarrow  \texttt{Dis}_o(\mathcal{H}_o(t))=\texttt{T}
\]
That is,  
\begin{align}
& (\forall t\in\mathcal{L}(G): \mathcal{H}_a(t)=\mathcal{H}_a(s)) \nonumber\\\
&(\forall  w_1,w_2  \in \mathcal{L}(G): \mathcal{H}_o(t)=\mathcal{H}_o(w_1)=\mathcal{H}_o(w_2)) \nonumber\\\
&[ (\delta(t), (\delta(w_1),\delta(w_2)))\notin X\times T_\texttt{spec} 
]\nonumber
\end{align}
Then by considering string $\mathcal{H}_a(s)$,  
we  know that for any$(x,(x_1,x_2))\in f_P(q_{0,P},\mathcal{H}_a(s))$, we have $ (x,(x_1,x_2))\notin X\times T_\texttt{spec}$. 
This means that $f_P(q_{0,P}, \mathcal{H}_a(s))\in Q_P^y$, i.e., $Q_P^y\ne\emptyset$.
\hfill $\qed$
\end{pf}

\bibliographystyle{plain}        
\bibliography{main}

\end{document}